\documentclass[aps,prfluids,preprint,groupedaddress,longbibliography]{revtex4-1}

\usepackage{graphicx}
\usepackage{newtxtext}
\usepackage{natbib}
\usepackage{hyperref}
\hypersetup{
    colorlinks = true,
    urlcolor   = blue,
    citecolor  = black,
}

\newcommand{\RomanNumeralCaps}[1]
\linenumbers

\usepackage{amsmath,amssymb,esint,bbm,textgreek,mathrsfs,mathtools}
\usepackage{setspace}
\usepackage{etoolbox}

\usepackage{stackengine}
\usepackage{scalerel}

\usepackage{cleveref}
\usepackage{algpseudocode}
\usepackage{algorithm}
\usepackage{caption}
\usepackage{subfigure}
\usepackage[normalem]{ulem}
\usepackage[table]{xcolor}
\usepackage{booktabs}
\usepackage{placeins}

\crefname{equation}{}{}

\graphicspath{{Figs2/}}
\usepackage{Matharu_commands}

\begin{document}


\title{Optimal Eddy Viscosity in Closure Models for 2D Turbulent
  Flows}


\author{Pritpal \surname{Matharu}}
\email[Email: ]{mathap1@mcmaster.ca}
\homepage[Website: ]{https://ms.mcmaster.ca/~mathap1/}
\author{Bartosz \surname{Protas}}
\email[Email: ]{bprotas@mcmaster.ca}
\homepage[Website: ]{https://ms.mcmaster.ca/~bprotas/}
\affiliation{Department of Mathematics and Statistics, McMaster University, Hamilton, ON L8S 4K1, Canada}


\date{\today}

\begin{abstract}
  We consider the question of fundamental limitations on the
  performance of eddy-viscosity closure models for turbulent flows,
  focusing on the Leith model for 2D {Large-Eddy Simulation}.  Optimal
  eddy viscosities depending on the magnitude of the vorticity
  gradient are determined subject to minimum assumptions by solving
  PDE-constrained optimization problems defined such that the
  corresponding optimal Large-Eddy Simulation best matches the
  {filtered Direct Numerical Simulation. First, we consider
    pointwise match in the physical space and the} main finding is
  that with a fixed cutoff wavenumber $k_c$, the performance of the
  Large-Eddy Simulation systematically improves as the regularization
  in the solution of the optimization problem is reduced and this is
  achieved with the optimal eddy viscosities exhibiting increasingly
  irregular behavior with rapid oscillations. Since the optimal eddy
  viscosities do not converge to a well-defined limit as the
  regularization vanishes, we conclude that {in this case} the
  problem of finding an optimal eddy viscosity does not in fact have a
  solution and is thus ill-posed. {We argue that this observation
    is consistent with the physical intuition concerning closure
    problems. The second problem we consider involves matching
    time-averaged vorticity spectra over small wavenumbers. It is
    shown to be better behaved and to produce physically reasonable
    optimal eddy viscosities. We conclude that while} better behaved
  and hence practically more useful eddy viscosities can be obtained
  with stronger regularization {or by matching quantities defined
    in a statistical sense,} the corresponding Large-Eddy Simulations
  will not achieve their theoretical performance limits.
\end{abstract}


\maketitle

\section{Introduction\label{sec:intro}}

The closure problem is arguably one of the most important outstanding
open problems in turbulence research. It touches upon some of the key
basic questions concerning turbulent flows and at the same time has
far-reaching consequences for many applications, most importantly, for
how we simulate turbulent flows in numerous geophysical, biological
and engineering settings. Given the extreme spatio-temporal complexity
of turbulent flows, accurate numerical solutions of the Navier-Stokes
system even at modest Reynolds numbers requires resolutions exceeding
the capability of commonly accessible computational resources. To get
around this difficulty, one usually relies on various simplified
versions of the Navier-Stokes system obtained through different forms
of averaging and/or filtering, such as the Reynolds-Averaged
Navier-Stokes (RANS) system and the Large-Eddy Simulation (LES).
However, such formulations are not closed, because these systems
involve nonlinear terms representing the effect of unresolved subgrid
stresses on the resolved variables. The ``closure problem'' thus
consists in expressing these quantities in terms of resolved variables
such that the RANS or LES system is closed.

In general, closure models in fluid mechanics are of two main types:
algebraic, where there is an algebraic relationship expressing the
subgrid stresses in terms of the resolved quantities, and
differential, where this relationship involves an additional
partial-differential equation (PDE) which needs to be solved together
with the RANS or LES system. Most classical models are usually
formulated based on some ad-hoc, albeit well-justified, physical
assumptions. There exists a vast body of literature concerning the
design, calibration and performance of such models in various
settings. Since it is impossible to offer an even cursory survey of
these studies here, we refer the reader to the well-known monographs
\citep{Lesieur1993book,pope2000turbulent,davidson2015turbulence} for
an overview of the subject. Recently, there has been a lot of activity
centered on learning new empirical closure models from data using
methods of machine learning
\citep{kutz_2017,GamaharaHattori2017,jimenez_2018,duraisamy2018turbulence,Duraisamy2021,San2021}.
It is however fair to say that the field of turbulence modelling has
been largely dominated by empiricism and there is a consensus that the
potential and limitations of even the most common models are still not
well understood. Our study tackles this fundamental question, more
specifically, how well certain common closure models can in principle
perform if they are calibrated in a optimal way.{ We will look
  for an optimal, in a mathematically precise sense, form of a certain
  closure model and will conclude that, somewhat surprisingly, it does
  not in fact exist.} 

{On the other hand, from the physical point of view, turbulence
  closure models are not meant to capture nonlinear transfer processes
  with pointwise accuracy, but rather to represent them in a certain
  average sense. The ill-posedness of the problem of optimally
  calibrating a closure model signalled above can thus be viewed as a
  consequence of the inability of the closure model to match the
  original solution pointwise in space and in time. More precisely,
  the optimal eddy viscosity exhibits unphysical high-frequency
  oscillations.  In the present study we will use a novel and
  mathematically systematic approach to illustrate this physical
  intuition and demonstrate how the ill-posedness arises. We will also
  show that the model calibration problem is in fact well-behaved when
  the LES with a closure model is required to match quantities defined
  in the statistical rather than pointwise sense.}


We are going to focus on an example from a class of widely used
algebraic closure models, namely, the Smagorinsky-type eddy-viscosity
models \citep{smagorinsky1963general} for LES. More specifically, we
will consider the Leith model \citep{Leith1968, Leith1971, Leith1996}
for two-dimensional (2D) turbulent flows. Like all eddy-viscosity
closure models, the Leith model depends on one key parameter which is
the eddy viscosity typically taken to be a function of some flow
variable.  Needless to say, performance of such models critically
depends on the form of this function. {One specific question we are
  interested in is how accurately the LES equipped with such an
  eddy-viscosity closure model can at best reproduce solutions of the
  Navier-Stokes system obtained via Direct Numerical Simulation (DNS).
  Another related question we will consider concerns reproducing
  certain statistical properties of Navier-Stokes flows in LES.  We
  will address these questions by formulating them as PDE-constrained
  optimization problems} where we will seek an optimal functional
dependence of the eddy viscosity on the state variable. {In the first
  problem we will require the corresponding LES to match the
  {filtered} DNS pointwise in space over a time window of several eddy
  turnover times, whereas in the second problem the LES will be
  required to match the time-averaged enstrophy spectrum of the
  Navier-Stokes flow for small wavenumbers. By framing these questions
  in terms of optimization problems we will be able to find the best
  (in a mathematically precise sense) eddy viscosities,} and this will
in turn allow us to establish ultimate performance limitations for
this class of closure models. We emphasize that the novelty of our
approach is that by finding an optimal functional form of the eddy
viscosity we identify, subject to minimum assumptions, an optimal
structure of the nonlinearity in the closure model, which is
fundamentally different, and arguably more involved, than calibrating
one or more constants in a selected ansatz for the eddy viscosity.
This formulation is also more general than common dynamic closure
models and some formulations employing machine learning to deduce
information about local properties of closure models from the DNS
(see, e.g., \citep{Maulik2020}). Our goal is to understand what form
the eddy viscosity needs to take in order to maximize the performance
of the closure model in achieving a prescribed objective. {The
  emphasis will be on methodology rather than on specific
  contributions to subgrid modeling.}

The optimization problem in question has a non-standard structure, but
an elegant solution can be obtained using a generalization of the
adjoint-based approach developed by
\citet{bukshtynov2011optimal,bukshtynov2013optimal}. {In being based
  on methods of the calculus of variations, this approach thus offers
  {a mathematically rigorous} alternative to machine-learning methods
  which have recently become popular
  \citep{kutz_2017,GamaharaHattori2017,jimenez_2018,duraisamy2018turbulence,Duraisamy2021,San2021}.}
As a proof of the concept applicable to the problem considered here,
this approach was recently adapted to find optimal closures in a
simple one-dimensional (1D) model problem by \citet{Matharu2020}.
Importantly, this approach involves a regularization parameter
controlling the ``smoothness'' of the obtained eddy viscosity.

{In the first problem, which involves matching the filtered DNS
  solution in the pointwise sense,} we find optimal eddy viscosities
for the Leith closure model in the LES systems with different filter
cutoff wavenumbers $k_c$. As this wavenumber increases and the filter
width vanishes, the optimal eddy viscosity is close to zero and the
match between the predictions of the LES and {the filtered} DNS
is nearly exact, {as expected}. On the other hand, for smaller cutoff
wavenumbers $k_c$ the optimal eddy viscosity becomes highly irregular
whereas the match between the LES and DNS deteriorates, although it
still remains much better than the match involving the LES with the
standard Leith model or with no closure model at all.  Interestingly,
the optimal eddy viscosity reveals highly oscillatory behavior with
alternating positive and negative values as the state variable
increases. When the regularization in the solution of the optimization
problems is reduced {and the numerical resolution is refined} at a
fixed cutoff wavenumber, the frequency and amplitude of these
oscillations are amplified which results in an improved match against
the DNS. {Thus, in this limit the optimal eddy viscosity becomes
  increasingly oscillatory as a function of the state variable which}
suggests that in the absence of regularization the problem of finding
an optimal eddy viscosity does not in fact have a solution as the
limiting eddy viscosity is not well defined. On the other hand, {an
  arbitrarily regular eddy viscosity} can be found when sufficient
regularization is used in the solution of the optimization problem,
{but at the price of reducing the match against the DNS. While such
  smooth eddy viscosities may be more useful in practice, the
  corresponding LES models will not achieve their theoretical
  performance limits.} In addition to this observation, our results
also demonstrate how the best accuracy achievable by the LES with the
considered closure model depends on the cutoff wavenumber of the
filter, which sheds new light on the fundamental performance
limitations inherent in this closure model.

{In our second problem, which involves matching the time-averaged
  vorticity spectrum of the filtered DNS, the obtained optimal eddy
  viscosity is more regular and its key features remain essentially
  unchanged as the regularization in the solution of the optimization
  problems is reduced and the numerical resolution is refined. This
  demonstrates that the problem of optimally calibrating the closure
  model is better behaved when a suitable statistical quantity is used
  as the target. This is not surprising as such a formulation is in
  fact closer to the spirit of turbulence modelling.}

The structure of the paper is as follows: in the next section we
formulate our LES model and state the optimization problem defining
the optimal eddy viscosity; in {Section} \ref{sec:adjoint} we
introduce an adjoint-based approach to the solution of the
optimization problem and in {Section} \ref{sec:numer} discuss
computational details; our results are presented in {Section}
\ref{sec:results} whereas final conclusions are deferred to {Section}
\ref{sec:final} {; some additional technical material is provided
  in {Appendix \ref{sec:gradJ2}}.}

\section{Large-Eddy Simulation and Optimal Eddy Viscosity\label{sec:LES}}

We consider 2D flows of viscous incompressible fluids on a periodic
domain $\Omega := [0,2\pi]^2$ over the time interval $[0,T]$ for some
$T> 0$ (``:='' means ``equal to by definition''). Assuming the fluid
is of uniform unit density $\rho = 1$, its motion is governed by the
Navier-Stokes system written here in the vorticity form
\begin{subequations} \label{eq:2DNS}
	\begin{alignat}{2} 
	\partial_t {w} + \gradperp \psi \cdot \bfgrad {w} &= \nuN \bfDelta {w}  - \a {w} + \fw & \quad &\text{in} \quad {(0, T] \times \Omega},  \label{eq:vort} \\
	\bfDelta \psi &= -{w} & \quad &\text{in} \quad {(0, T] \times \Omega}, \label{eq:stream_vort} \\
	{w}(t=0) &= {w_0} & \quad &\text{in} \quad \Omega, \label{eq:IC}
	\end{alignat} 
\end{subequations}
where {$w = -\gradperp \cdot \u$}, with $\gradperp =
[\partial_{x_2}, -\partial_{x_1}]^{T}$ and $\u$ the velocity field, is
the vorticity component perpendicular to the plane of motion, $\psi$
is the streamfunction, $\nuN$ is the coefficient of the kinematic
viscosity (for simplicity, we reserve the symbol $\nu$ for the eddy
viscosity), and ${w_0}$ is the initial condition.  System
\eqref{eq:2DNS} is subject to two forcing mechanisms: a
time-independent forcing $\fw$ which ensures that the flow remains in
a statistical equilibrium and the {Ekman} friction $-\alpha {w}$
describing large-scale dissipation due to, for example, interactions
with boundary layers arising in geophysical fluid phenomena. The
forcing term is defined to act on Fourier components of the solution
with wavenumbers in the range $[k_a,k_b]$ for some $0 < k_a < k_b <
\infty$, i.e.,
\begin{align} \label{eq:f}
\left[\widehat{f}_{\w}\right]_{\kb} := 
\begin{cases}
F, \quad &k_a \leq |\kb| \leq k_b, \\
0, \quad &\text{otherwise},
\end{cases}
\end{align}
where $\left[\widehat{f}_{\w}\right]_{\kb}$ is the Fourier component
of $\fw$ with the wavevector $\kb$ (hereafter hats
``\;$\widehat{\cdot}$\;'' will denote Fourier coefficients) and $F>0$
is a constant parameter.

The phenomenology of 2D forced turbulence is described by the
Kraichnan-Batchelor-Leith theory
\citep{Kraichnan1967,Batchelor1969,Leith1968} which makes predictions
about various physical characteristics of such flows. Their prominent
feature, distinct from turbulent flows in three dimensions (3D), is
the presence of a forward enstrophy cascade and an inverse energy
cascade
\citep{Boffetta2007,Boffetta2010,Bracco2010,Vallgren2011,Boffetta2012}.
Here we will chose $k_a$ and $k_b$ such that the forcing term
\eqref{eq:f} will act on a narrow band of Fourier coefficients to
produce a well-developed enstrophy cascade towards large wavenumbers
and a rudimentary energy cascade towards small wavenumbers. The
parameters $\nuN$, $\alpha$ and $F$ will be adjusted to yield a
statistically steady state with enstrophy $\E(t) := \intO {w}^2(t,\x)
\, d\Omega$ fluctuating around a well-defined mean value $\E_0$. {The
  initial condition $\omega_0$ in \cref{eq:IC} will be chosen such
  that the evolution begins already in this statistically steady state
  at time $t=0$.}

\subsection{The Leith Closure Model}
\label{sec:Leith}

The LES is obtained by applying a suitable low-pass filter
$G_{\delta}$, where $\delta>0$ is its width, to the Navier-Stokes
system \eqref{eq:2DNS} and defining the filtered variables
${\widetilde{w}} = G_{\delta} \ast {w}$ and $\tpsi = G_{\delta} \ast
\psi$ (``\;$\ast$\;'' denotes the convolution operation and hereafter
we will use tilde ``\;$\widetilde{\cdot}$\;'' to represent filtered
variables). For simplicity, we will employ a sharp {low-pass
  spectral} filter defined in terms of its Fourier-space
representation as
\begin{equation} 
\label{eq:G}
\left[\widehat{G}_{\delta}\right]_{\kb} :=
\begin{cases}
1, &|\kb| \leq {k_c}, \\
0, &\text{otherwise},
\end{cases}
\end{equation}
where ${k_c}$ is the largest resolved wavenumber such that the filter
width is $\delta = 2\pi / {k_c}$. Since we normally have $k_b <
{k_c}$, it follows that $\tfw = \fw$. Application of filter
\eqref{eq:G} to the vorticity equation \eqref{eq:vort} yields $\partial_t {\widetilde{w}} + {\superwidetilde{\gradperp \tpsi \cdot \bfgrad {\widetilde{w}}}} = \nuN
\bfDelta {\widetilde{w}} - \a {\widetilde{w}} + \fw + M$, where the term $M$ represents the
effect of the unresolved subgrid quantities
\begin{equation} 
M = {\superwidetilde{\gradperp \tpsi \cdot \bfgrad {\widetilde{w}}} - \superwidetilde{\gradperp \psi \cdot \bfgrad {w}}}. 
\label{eq:M}
\end{equation}
Since expression \eqref{eq:M} depends on the original unfiltered
variables ${w}$ and $\psi$, to close the filtered system the term $M$
must be modelled in terms of an expression involving the filtered
variables only. We will do this using the {Leith} model
\citep{Leith1968, Leith1971, Leith1996}{,} which has a similar
structure to the Smagorinsky model \citep{smagorinsky1963general}
widely used as a closure for 3D flows, but is derived considering the
forward enstrophy cascade as the dominant mechanism in 2D turbulent
flows. There is evidence {for good performance of the {Leith}
  model in} such flows \citep{Graham2013, Maulik2017}. {Its
  preferred form is}
\begin{equation} 
\label{eq:Leith}
M \approx \widetilde{M} = \bfgrad \cdot (\widetilde{{\nu_L} \gtw}), 
\end{equation}
{in which} {$\tw$ is the solution to the LES system,
  cf.~\cref{eq:LES}, and} the eddy viscosity is assumed to be a {linear
function of the magnitude of the vorticity gradient, i.e., 
\begin{equation}
\nu_L(s) := (C_L \delta)^3 \, \sqrt{s}  \qquad \text{with} \quad s :=
|\gtw|^2 \in \I := [0,\smax],
\label{eq:nu0}
\end{equation}
where the Leith constant $C_L = 1$ and} $\smax > 0$ is a sufficiently
large number to be specified later. We will refer to $\I$ as the
{``state space'' domain}.

{While in the original formulation of the Leith model the eddy
viscosity is taken to be a linear function of $|\gtw|$ as  in \eqref{eq:nu0} 
\citep{Graham2013, Maulik2017}, here we consider a general
dependence of the eddy viscosity on $|\gtw|$ in the form
\begin{equation}
  \nu(s) = \left[ \nu_L(s) + \nu_0 \right] \varphi\left( \frac{s}{\smax} \right),
  \label{eq:nu}
\end{equation}
where $\nu_0 > 0$ and $\varphi \; : \; [0,1] \rightarrow \RR$ is a
dimensionless function subject to some minimum only assumptions to be
specified below. The parameter $\nu_0$ is introduced to allow the eddy
viscosity $\nu(s)$ to take nonzero values at $s = 0$, in contrast to
Leith's original model \eqref{eq:nu0}. We remark that defining the
eddy viscosity in terms of such a function $\varphi$ ensures that
ansatz \eqref{eq:nu} is dimensionally consistent. Making $\varphi$ and
$\nu$ functions} of $|\gtw|^2$, rather than of $|\gtw|$, in
\eqref{eq:nu} will simplify subsequent calculations. {We add that
  ansatz \eqref{eq:nu} is used here to illustrate the approach and in
  principle one could also consider other formulations parametrized by
  nondimensional functions.}  With the Leith model
\eqref{eq:Leith}--\eqref{eq:nu}, the LES version of the 2D
Navier-Stokes system \eqref{eq:2DNS} takes the form
\begin{subequations} \label{eq:LES}
	\begin{alignat}{2} 
	\partial_t \tw + {\superwidetilde{\gradperp \tpsi \cdot \bfgrad \tw}} &= \bfgrad \cdot {\ssuperwidetilde{\left( \left[\nuN + \nu(s) \right] \bfgrad \tw \right)}}  - \a \tw + \fw & \quad &\text{in} \quad {(0, T] \times \Omega},  \label{eq:LES_eqn} \\
	\bfDelta \tpsi &= -\tw & \quad &\text{in} \quad {(0, T] \times \Omega}, \label{eq:LES_stream_vort} \\
	\tw(t=0) &= \tw_0 {:=} \widetilde{w}_0& \quad &\text{in} \quad \Omega, \label{eq:LES_IC}
	\end{alignat} 
\end{subequations}
where the initial condition is given as the filtered initial condition
\eqref{eq:IC} from the DNS system.

An equivalent form of equation \eqref{eq:LES_eqn} can be obtained
noting that with the form of the filter given in \eqref{eq:G}, the
{decomposition} of the subgrid stresses \eqref{eq:M} reduces to $M =
\gradperp \tpsi \cdot \bfgrad \tw - {\superwidetilde{\gradperp \psi
    \cdot \bfgrad \w}}$ \citep{pope2000turbulent}. As a result, the
advection term in \eqref{eq:LES_eqn} can be replaced with $\gradperp
\tpsi \cdot \bfgrad \tw$. While our numerical solution will be based
on \eqref{eq:LES_eqn}, this second form will facilitate the
derivations presented in {Section} \ref{sec:adjoint}. We will assume that
for all times $t \in [0,T]$ the filtered vorticity field $\tw$ is in
the Sobolev space {$H_{0}^2(\Omega)$} of {zero-mean} functions
with square-integrable second derivatives \citep{af05}. {We
  stress the distinction between the fields $w$, $\widetilde{w}$,
  $\tw$ which represent, respectively, the solution of the DNS system
  \cref{eq:2DNS}, its filtered version and the solution of the LES
  system \cref{eq:LES}.}

\subsection{Optimization Formulation for Eddy Viscosity}
\label{sec:opt}

{We consider two formulations with the DNS field matched
  pointwise in space and in time, and in a certain statistical sense.
  First, the} optimal eddy viscosity will be {found} as a minimizer of
an error functional representing the mean-square error between
{observations of the filtered DNS, i.e., of the filtered solution
  $\widetilde{w}(t,\x)$ of the Navier-Stokes system \eqref{eq:2DNS},
  and observations} the corresponding prediction
$\tw(t,\x;{\varphi})$ of the LES model \eqref{eq:LES} with eddy
viscosity $\nu$.  These observations are acquired at points $\x_i$,
$i=1,\dots,M^2$, forming a uniform $M \times M$ grid in $\Omega$ with
operators $H_{i} \; : \; {H^2}(\Omega) \longrightarrow \mathbb{R}$
defined as
\begin{equation}
\left(H_i {\tw}\right)(t) := \intO \delta(\x - \x_i) {\tw}(t, \x) \, d\Omega = {\tw}(t,\x_i), 
\quad i=1,\dots,M^2,
\label{eq:Hi}
\end{equation}
where $\delta(\cdot)$ is the Dirac delta distribution and observations
$\left(H_i \tw({\varphi})\right)(t)$ of the LES solution are defined analogously
(an integral representation of the observation operators will be
convenient for the derivation of the solution approach for the
optimization problem presented in {Section} \ref{sec:adjoint}). The number
of the observations points $M^2$ will be chosen such that $M
\gtrapprox k_c$, i.e., the observations will resolve all flow features
with wavenumbers slightly higher than the cutoff wavenumber $k_c$ in
\eqref{eq:G}. The error functional then takes the form 
\begin{equation} 
\label{eq:J}
{\J_1}({{\varphi}}) := \frac{1}{2} \int_0^T \sum_{i = 1}^{M^2} \left[\left(H_i {\widetilde{w}}\right)(t)  - \left(H_i \tw({\varphi})\right)(t) \right]^2 \, dt,
\end{equation}
and is understood as depending on the {function $\varphi$
  parametrizing the eddy viscosity $\nu = \nu(s)$ via ansatz
  \eqref{eq:nu}.}

{In the second formulation, the optimal eddy viscosity will be
  found by minimizing the error between the time-averaged vorticity
  spectra in the filtered DNS and predicted by the LES. For
  simplicity and with a slight abuse of notation, we will treat the
  wavenumber $k$ as a continuous variable, i.e., we will assume that
  $\kb \in \RR^2$ rather than $\kb \in \ZZ^2$; in the actual
  implementation one needs to account for the discrete nature of the
  wavevector $\kb$. The vorticity spectrum predicted by the LES is
  then defined as
\begin{align}
\label{eq:E}
E_{\tw}(t, k) := \frac{1}{2} \int_{\mathscr{C}(k)} |\htw(t, \kb)|^2 \, dS(\kb),
\qquad \forall t, k \ge 0,
\end{align}
where $\htw(t, \kb)$ is the Fourier transform of $\tw(t,\x)$ and
$\mathscr{C}(k) := \{\kb \in \RR^2 : |\kb|=k\}$ a circle with radius
$k$ in the $2$D plane.  The vorticity spectrum $E_w(t,k)$ in the
(filtered) DNS is defined analogously.  Denoting
$[f]_T := (1/T) \int_0^T f(t)\, dt$ the time average of a function
$f \: : \; [0,T] \rightarrow \RR$, the error functional is defined as
\begin{equation} 
\label{eq:J2}
{\J_2(\varphi) := \frac{1}{4} \int_{k = 0}^{k_c} \left( \left[ \Ew(\cdot, k; \varphi) \right]_T -  \left[ E_w(\cdot, k) \right]_T  \right)^2 \, dk,}
\end{equation}
{with matching performed up to the cutoff wavenumber $k_c$.}
}

The form of equation \eqref{eq:LES_eqn} suggests that $\nu = \nu(s)$,
{and hence also $\varphi = \varphi( s / \smax )$, must be at
  least piecewise $C^1$ functions on $\I$ and $[0,1]$, respectively.}
However, as will become evident in {Section} \ref{sec:adjoint}, our
solution approach imposes some additional regularity requirements,
namely, $\nu = \nu(s)$ needs to be piecewise $C^2$ on $\I$ with the
first and third derivatives vanishing at $s = 0,\smax$. Since
gradient-based solution approaches to PDE-constrained optimization
problems are preferably formulated in Hilbert spaces \citep{pbh04}, we
shall look for an optimal {function $\varphi$ parametrizing the
  optimal} eddy viscosity as an element of the following linear space
which is a subspace of the Sobolev space $H^2(\I)$
\begin{equation}
\cS := {\left\{ \varphi \in {C^3([0,1])} \: : \: \frac{d}{d\xi} \varphi(\xi) =  \frac{d^3}{d\xi^3} \varphi(\xi) = 0 \ \text{at} \ \xi = 0,1 \right\}}.
\label{eq:S}
\end{equation}
Then, the problem of finding an optimal eddy viscosity {in the
  two formulations becomes 
\begin{equation}
\label{eq:minJ}
\varphic:= \underset{\varphi \in \cS} {\argmin} \, \J_j(\varphi), \qquad j = 1, 2,
\end{equation}
where the optimal eddy viscosity $\nuc$ is deduced from $\varphic$ via
ansatz \eqref{eq:nu}.}  Our approach to solving this problem is
outlined in the next section.

\section{Adjoint-based Optimization} 
\label{sec:adjoint}

{To fix attention, we focus here on solution of the optimization
  problem in the first formulation, i.e., for $j=1$ in
  \eqref{eq:minJ}, with the error functional given in
  \eqref{eq:J}. Essentially the same approach can also be used to
  solve the second optimization problem with the error functional
  \eqref{eq:J2} and required modifications are discussed in
  \Cref{sec:gradJ2}.}  We formulate our approach in the continuous
(``optimize-then-discretize'') setting \citep{g03} and adopt the
strategy developed and validated by \citet{Matharu2020}. Here we only
summarize its key steps and refer the reader to that study for further
details.  A local solution of problem {\cref{eq:J},
  \cref{eq:S}--\cref{eq:minJ}} can be found using an iterative
gradient-based minimization approach as
{$\varphic = \underset{n \to \infty}{\text{lim}} \varphi^{(n)}$,
  where
\begin{align} 
\label{eq:desc}
\begin{cases}
\varphi^{(n+1)} &= \varphi^{(n)} - \tau^{(n)} \, \grad_{\varphi}\J_1(\varphi^{(n)}), \quad \quad n=0, 1, \dots, \\
\varphi^{(0)} &= \varphi_0,
\end{cases}
\end{align}
in which $\varphi^{(n)}$ is the approximation of the optimal function
$\varphic$ at the $n$th iteration (which can be used to construct the
corresponding approximation $\nu^{(n)}$ of the optimal eddy
viscosity), {$\grad_{\varphi}\J_1(\varphi)$} is the gradient of the
error functional \eqref{eq:J} with respect to $\varphi$, $\tau^{(n)}$
is the step length along the descent direction and $\varphi_0$ is an
initial guess usually suggested by some form of the eddy viscosity.}

A central element of algorithm \eqref{eq:desc} is the gradient
{$\grad_{\varphi}{\J_1}(\varphi)$}. In many problems of
PDE-constrained optimization it can be conveniently expressed using
solutions of suitably-defined adjoint equations \citep{g03}. However,
the present optimization problem {\cref{eq:J},
  \cref{eq:S}--\cref{eq:minJ}} has a nonstandard structure because the
control variable {$\varphi(s/\smax)$} is a function of the {\em
  dependent} variable $s = |\gtw|^2$ in system \eqref{eq:LES}. On the
other hand, in its standard formulation adjoint analysis allows one to
obtain expressions for gradients depending on the {\em independent}
variables in the problem (here, $t$ and $\x$).  This difficulty was
overcome by \citet{bukshtynov2011optimal,bukshtynov2013optimal} who
generalized adjoint analysis of PDE systems to problems of the type
{\cref{eq:J}, \cref{eq:S}--\cref{eq:minJ}} by introducing a
suitable change of variables. {For convenience we will denote
  $\sigma:= s/\smax$.}

The G\^{a}teaux (directional) differential of the error functional
\eqref{eq:J} with respect to ${{\varphi}}$, defined by ${\J_1}'({{\varphi}}; {\varphi}') :=
\lim_{\epsilon\to0} \epsilon^{-1} \left[{\J_1}({{\varphi}} + \epsilon {\varphi}') -
  {\J_1}({{\varphi}})\right]$, is defined as
\begin{align} 
\label{eq:dJ}
{\J_1}'({{\varphi}}; {\varphi}')=& \int_0^T \intO \, \sum_{i = 1}^{M^2} H_{i}^{*}
\left[ \left(H_i \tw({\varphi}) \right)(t)  - \left(H_i {\widetilde{w}} \right)(t) \right] \, {\tw'}(t, \x; {{\varphi}}, {\varphi}') \, d\x \, dt,
\end{align}
where ${\varphi}' \in \cS$ is an arbitrary perturbation of the
{control variable} ${\varphi}$, $\tw'(t, \x;
{{\varphi}}, {\varphi}')$ satisfies the system {
\begin{subequations}
	\label{eq:Pert}
	\begin{align}
	&\K\begin{bmatrix} \tw' \\  \\ \tpsi' \end{bmatrix} := 
	 \begin{bmatrix}
	\partial_t \tw' + \gradperp \tpsi' \cdot \gtw + \gradperp \tpsi \cdot \gtw' + \a \tw' \qquad \qquad \qquad \\ 
	- \bfgrad \cdot \left(2 (\gtw \cdot \gtw') (\frac{d \nu}{d{s}} \, \varphi \gtw + \frac{\nu_L + \nu_0}{\smax}\, \frac{d \varphi}{d\sigma} \, {\gtw} )+ (\nuN + \nu) \gtw' \right)\\
	\bfDelta \tpsi' + \tw' 
	\end{bmatrix} = \begin{bmatrix} \bfgrad \cdot \left( (\nu_L + \nu_0) {\varphi}'  \gtw \right) \\  \\ 0\end{bmatrix}, \label{eq:Pert_eqn} \\
	&\tw'({t=0, \x}) = 0, \label{eq:Pert_IC}
	\end{align}
\end{subequations}}
obtained as linearization of the LES system \cref{eq:LES} and $H_{i}^*
\; : \; \mathbb{R} \longrightarrow {H^{-2}}(\Omega)$, $i = 1, \dots,
M^2$, are the adjoints of the observation operators $H_{i}$,
{cf.~\cref{eq:Hi}, given by}
\begin{equation}
  {\forall \xi \in \RR, \qquad \left(H_i^{*} \xi\right) := \delta(\x - \x_i) \xi,} \quad i=1,\dots,M^2.
\end{equation}
In order to extract the gradient $\grad_{{\varphi}}{\J_1}$ from the G\^{a}teaux
differential \eqref{eq:dJ}, we note that this derivative is a bounded
linear functional when viewed as a function of ${\varphi}'$ and invoke the
Riesz representation theorem \citep{b77} to obtain
\begin{equation} 
\label{eq:Riesz}
{\J_1}'({{\varphi}}; {\varphi}') = \left\langle \grad_{{\varphi}}^{H^2}{\J_1}, {\varphi}' \right\rangle_{H^2({[0, 1]})} = \left\langle \grad_{{\varphi}}^{L^2}{\J_1}, {\varphi}' \right\rangle_{L^2({[0, 1]})},
\end{equation}
where the inner product in the space $H^2({[0, 1]})$ is defined as
{
\begin{equation}
\Big\langle p_1, p_2 \Big\rangle_{H^2({[0, 1]})} 
 = \int_0^1 p_1 \, p_2 + \ell_1^2 \, \frac{dp_1}{d\sigma}\frac{dp_2}{d\sigma} + \ell_2^4 \, \frac{d^2p_1}{d\sigma^2}\frac{d^2p_2}{d\sigma^2}  \, d\sigma, 
\label{eq:ipH2}
\end{equation}}
in which $\ell_1$ and $\ell_2$ are length-scale parameters. While for
all values of $\ell_1,\ell_2 \in (0,\infty)$ the inner products
\eqref{eq:ipH2} are equivalent (in the sense of norm equivalence),
these two parameters play a very important role in regularization of
solutions to the optimization problem \eqref{eq:J}--\eqref{eq:minJ}.
In \cref{eq:desc} we require the gradient in the space $H^2({[0, 1]})$, i.e.,
$\grad_{{\varphi}}{\J_1} = \grad_{{\varphi}}^{H^2}{\J_1}$, but it is convenient to first
derive the gradient with respect to the $L^2$ topology.

Introducing {\em adjoint fields} $\tw^*$ and $\tpsi^*$, {we can
  define the following duality-pairing relation
\begin{equation}
\begin{aligned}
\left( \K\begin{bmatrix} \tw' \\ \tpsi' \end{bmatrix}, \begin{bmatrix} \tw^* \\ \psi^* \end{bmatrix} \right)
:= & \int_0^T \intO \K\begin{bmatrix} \tw' \\ \tpsi' \end{bmatrix} \cdot \begin{bmatrix} \tw^* \\ \tpsi^* \end{bmatrix} \, d\x \, dt \\
& \int_0^T \intO \begin{bmatrix} \tw' \\ \tpsi' \end{bmatrix} \cdot \K^* \begin{bmatrix} \tw^* \\ \tpsi^* \end{bmatrix} \, d\x \, dt = \left( \begin{bmatrix} \tw' \\ \tpsi' \end{bmatrix}, \K^*\begin{bmatrix} \tw^* \\ \psi^* \end{bmatrix} \right),
\end{aligned}
\label{eq:dual}
\end{equation}
where integration by parts was performed with respect to both space
and time (noting the periodic boundary conditions and the initial
condition \eqref{eq:Pert_IC}) and the {\em adjoint system} has the
form}
{
\begin{subequations}
	\label{eq:Adj}
	\begin{align}
	&\K^*\begin{bmatrix} \tw^* \\  \\ \tpsi^* \end{bmatrix} := 
	 \begin{bmatrix}
	-\partial_t \tw^* - \gradperp \tpsi \cdot \gtw^* + \a \tw^* + \tpsi^* \qquad \qquad \qquad \qquad \\ 
	- \bfgrad \cdot \left(2 \, (\gtw \cdot \gtw^*) \, (\frac{d \nu}{d{s}} \, \varphi \gtw + \frac{\nu_L + \nu_0}{\smax}\, \frac{d \varphi}{d\sigma} \, {\gtw} ) + (\nuN + \nu) \gtw^*\right)\\ 
	\bfDelta \tpsi^* - \gradperp \cdot (\tw^* \, \gtw)
	\end{bmatrix} = \begin{bmatrix}  W \\  \\ 0\end{bmatrix}, \label{eq:Adj_eqn} \\
	&\tw^*({t=T, \x}) = 0, \label{eq:Adj_IC}
	\end{align}
\end{subequations}}
with the source term $W(t,\x) := \sum_{i = 1}^{M^2}
H_i^*\left[\left(H_i \tw({\varphi}) \right)(t) - \left(H_i {\widetilde{w}}
  \right)(t)\right]$. {Combining \eqref{eq:Pert}, \eqref{eq:dual}
  and \eqref{eq:Adj}, we then arrive at
\begin{equation}
\begin{aligned}
\left( \begin{bmatrix} \tw' \\ \tpsi' \end{bmatrix}, \K^*\begin{bmatrix} \tw^* \\ \tpsi^* \end{bmatrix}\right) 
& = \overbrace{\int_0^T \intO \, {W(t, \x)} \, \tw' \, d\x dt}^{{\J_1}'({{\varphi}}; {\varphi}')} \\
& =  {-\int_0^T \intO \, \left(\nu_L + \nu_0 \right) \, \left(\gtw \cdot \gtw^* \right) \, {\varphi}' \, d\x \, dt, }
\end{aligned}
\label{eq:dual_Adj}
\end{equation}
from which we obtain an expression for the G\^{a}teaux differential 
\begin{align*}
{
{\J_1}'({{\varphi}}; {\varphi}') = -\int_0^T \intO \, \left(\nu_L + \nu_0 \right) \, \left(\gtw \cdot \gtw^*\right)
\, {\varphi}' \, d\x \, dt{,}
}
\end{align*}
with the perturbation ${\varphi}'$ now appearing
explicitly as a factor. However, this expression is still not
consistent with the Riesz form \eqref{eq:Riesz}, which requires
integration with respect to $s$ over {$[0,1]$}.  In order to perform the 
required change of variables, we make the substitution
{$\varphi'(\gtw \cdot \gtw) = \int_0^1 \, \delta\left(\frac{\gtw \cdot \gtw}{\smax} - \sigma
\right) \, \varphi'(\sigma) \, d\sigma$}. Fubini's theorem then allows us to swap the
order of integration such that the G\^{a}teaux differential
\eqref{eq:dJ} is finally recast in the Riesz form \eqref{eq:Riesz} as
an integral with respect to {$\sigma$
\begin{equation} 
\label{eq:dJ2}
{\J_1}'({\varphi}; \varphi') = \int_0^1 \left[-\int_0^T \intO \, \delta\left(\frac{\gtw \cdot \gtw}{\smax} - \sigma \right) \, \left(\nu_L + \nu_0 \right) \,\gtw \cdot \gtw^* \, d\x \, dt \right] \,  \varphi'(\sigma) \, d\sigma.
\end{equation}}
{The gradient defined with respect to the $L^2$ topology is then
  deduced from this expression as}
{\begin{equation} 
\label{eq:gradL2}
\grad_{\varphi}^{L^2}{\J_1}(\sigma) 
=-\int_0^T \intO \, \delta\left(\frac{\gtw \cdot \gtw}{\smax} - \sigma \right) \, \left(\nu_L + \nu_0 \right) \, \gtw \cdot \gtw^* \, d\x \, dt.
\end{equation}}

The $L^2$ gradient given in \eqref{eq:gradL2} may in principle be
discontinuous as a function of $s$ and hence will not ensure the
regularity required of the optimal eddy viscosity, cf.~{Section}
\ref{sec:opt}. To circumvent this problem, we define a Sobolev
gradient using the Riesz relations \eqref{eq:Riesz} to identify the
$H^2$ inner product \eqref{eq:ipH2} with expression \eqref{eq:dJ2} for
the G\^{a}teaux differential.  Integrating by parts with respect to
{$\sigma$} and noting that the perturbation
${\varphi}' \in \cS$ is arbitrary, we obtain the Sobolev gradient
{$\grad_{{\varphi}}^{H^2}{\J}$} as a solution of the
elliptic boundary-value problem {\begin{subequations}
	\label{eq:gradH2BVP}
	\begin{align}
	\left[\Id - \ell_1^2 \, \frac{d^2}{d\sigma^2} + \ell_2^4 \frac{d^4}{d\sigma^4}\right] \grad_{\varphi}^{H^2}\J_1(\sigma) &= \grad_{\varphi}^{L^2}\J_1(\sigma), \qquad \sigma \in [0, 1], \label{eq:gradH2} \\
	\frac{d^{(1)} \, (\grad_{\varphi}^{H^2}\J_1)}{d\sigma^{(1)}} \Big|_{\sigma=0,1} &= \frac{d^{(3)} \, (\grad_{\varphi}^{H^2}\J_1)}{d\sigma^{(3)}} \Big|_{\sigma=0,1} = 0. \label{eq:gradH2bc}
	\end{align}
      \end{subequations}} 
    The choice of the boundary conditions in \eqref{eq:gradH2bc}
    ensures the vanishing of all the boundary terms resulting from the
    integration by parts. There is in fact some freedom in how to
    cancel these terms and the choice in \eqref{eq:gradH2bc} is
    arguably the least restrictive. {As argued in Section
      \ref{sec:Leith}, we allow the eddy viscosity $\nu(s)$ to take
      nonzero values at $s=0$ so the corresponding Sobolev gradient
      should not vanish at $\sigma = 0$ such that it can modify the
      value of $\varphi(0)$, which turns out to be important in
      practice, cf.~Section \ref{sec:results}. Thus, the choice of
      boundary conditions at $\sigma = 0$ provided in
      \eqref{eq:gradH2bc} is necessary.}  On the other hand, the
    choice of the boundary conditions at {$\sigma=1$} has been
    found to have little effect on the gradient and on the obtained
    results provided $\smax$ is sufficiently large. Therefore, the
    form of these boundary conditions given in \eqref{eq:gradH2bc} is
    justified by simplicity.}  The boundary conditions
  \eqref{eq:gradH2bc} are the reason for the presence of additional
  constraints in the definition of space $\cS$ in \eqref{eq:S}.

Determination of the Sobolev {gradients $\grad_{{\varphi}}^{H^2}\J_1$ based on
the $L^2$ gradients $\grad_{{\varphi}}^{L^2}{\J_1}$} by solving system
\eqref{eq:gradH2BVP} can be viewed as low-pass filtering of the latter
gradient using a non-sharp filter (as discussed by \citet{pbh04}, this
can be seen representing the operator $\left[\Id - \ell_1^2 \,
  (d^2/d{\sigma}^2) + \ell_2^4 (d^4/d{\sigma}^4)\right]^{-1}$ in the Fourier space).
The parameters $\ell_1$ and $\ell_2$ serve as cutoff length scales
representing the wavelengths of the finest features retained in the
{gradients $\grad_{{\varphi}}^{H^2}{\J_1}$} such that increasing $\ell_1$ and
$\ell_2$ has the effect of making the Sobolev gradient ``smoother''
and vice versa. Thus, $\ell_1$ and $\ell_2$ are ``knobs'' which can be
tuned to control the regularity of the optimal eddy viscosities
obtained as solutions of the problem \eqref{eq:J}--\eqref{eq:minJ}.

Since by construction $\grad_{{\varphi}}^{H^2}{\J_1} \in \cS$, choosing the
initial guess in \eqref{eq:desc} such that ${\varphi}_0 \in \cS$ will ensure
that ${\varphi}^{(0)},{\varphi}^{(1)},\dots,{\varphic} \in \cS$. At each step in
\eqref{eq:desc} an optimal step size $\tau^{(n)}$ can be found by
solving the following line-minimization problem \citep{nw00}
\begin{equation}
\tau^{(n)} = \argmin_{\tau > 0} {\J_1}({\varphi}^{(n)} - \tau \, \grad_{{\varphi}}{\J_1}({\varphi}^{(n)})).
\label{eq:taun}
\end{equation}
Numerical implementation of the approach outlined above is discussed
in the next section.

\section{Computational Approach \label{sec:numer}}

The evaluation of the Sobolev gradient
$\grad_{{\varphi}}^{H^2}{\J_1}$ requires the numerical
solutions of the LES system \cref{eq:LES} and the adjoint system
\cref{eq:Adj} followed by the solution of problem \cref{eq:gradH2BVP}.
For the first two {systems} we use a standard Fourier pseudo-spectral
method in combination with a CN/RKW3 time-stepping technique
{introduced by \citet{Moin1991}} which give spectrally accurate
results in space and a globally second-order accuracy in time. The
spatial domain is discretized using $N_x = 256$ equispaced grid points
in each direction. Since the eddy viscosity $\nu = \nu(s)$ {and
  the function $\varphi(s/\smax)$ are} state-dependent, we also need
to discretize the state domain $\I$, cf.~\eqref{eq:nu}, which is done
using $N_s$ Chebyshev points (values {of $N_s$ are provided} in Table
\ref{table:1}). We use Chebyshev differentiation matrices to perform
differentiation with respect to $s$ and the eddy viscosity $\nu(s)$
and its derivatives are interpolated from state space $\I$ to the
spatial domain $\Omega$ using the barycentric formulas
\citep{trefethen2013approximation}. The boundary-value problem
\eqref{eq:gradH2BVP} is solved using a method based on ultraspherical
polynomials available in the {\tt chebop} feature of Chebfun
\citep{driscoll2014chebfun}.  Solution of the 2D Navier-Stokes system
\cref{eq:2DNS} is dealiased using {Gaussian filtering based on the
  $3/2$ rule} \citep{hou2009blow}, however, this is unnecessary for
the LES system \cref{eq:LES} due to the aggressive filtering applied.
To ensure that aliasing errors resulting from the presence of
state-dependent viscosity are eliminated, the adjoint system
\cref{eq:Adj} is solved using twice as many grid points $2N_x$ in each
direction.

Evaluation of the $L^2$ gradient \cref{eq:gradL2} requires
non-standard integration over level sets as described by
\citet{bukshtynov2013optimal}. While for simplicity a simple gradient
approach was presented in \eqref{eq:desc}, in practice we use the
Polak-Ribi{\`{e}}re variant of the conjugate-gradient method to
accelerate convergence. For the line minimization problem
\cref{eq:taun}, the standard Brent's algorithm is used \citep{pftv86}.
The consistency and accuracy of the formulation and of the entire
computational approach was validated using a standard suite of tests
as was done by \citet{Matharu2020}.

\section{Results \label{sec:results}}

{The results obtained by solving optimization problem
  \eqref{eq:minJ} with error functionals \cref{eq:J} and \cref{eq:J2}
  are presented in Sections \ref{sec:resultsJ1} and
  \ref{sec:resultsJ2} below.}  Our computations are based on a flow
problem defined by the following parameters
$\nu_N = 1 \times 10^{-2}$, $\alpha = 1 \times 10^{-3}$, $F = 5$, and
$k_{a} = k_{b} = 4$. In the {first optimization problem we fix
  $M = 32$ in \eqref{eq:J}}, which is slightly larger than the largest
cutoff wavenumber $k_c$ we consider (cf.~Table \ref{table:1}) and
therefore ensures that the optimal eddy viscosity is determined based
on all available flow information, and $T = 20 \approx 30 t_e$, where
$t_e := \left[ \int_{0}^{T} \E(t) \, dt / ( 8\pi^2 T)\right]^{-1/2}$
is the eddy turnover time \citep{Bracco2010}. We emphasize that the
key insights provided by our computations do not depend on the
particular choice of $T$, as long as it remains of comparable
magnitude to the value given above.

{\subsection{Matching the DNS Pointwise in Space and Time ---
    Results for the Optimization Problem with Error Functional
    \eqref{eq:J} \label{sec:resultsJ1}}}

\begin{figure}\centering
  \subfigure[]
  {\hspace*{-1.75cm}
    \includegraphics[scale=0.6]{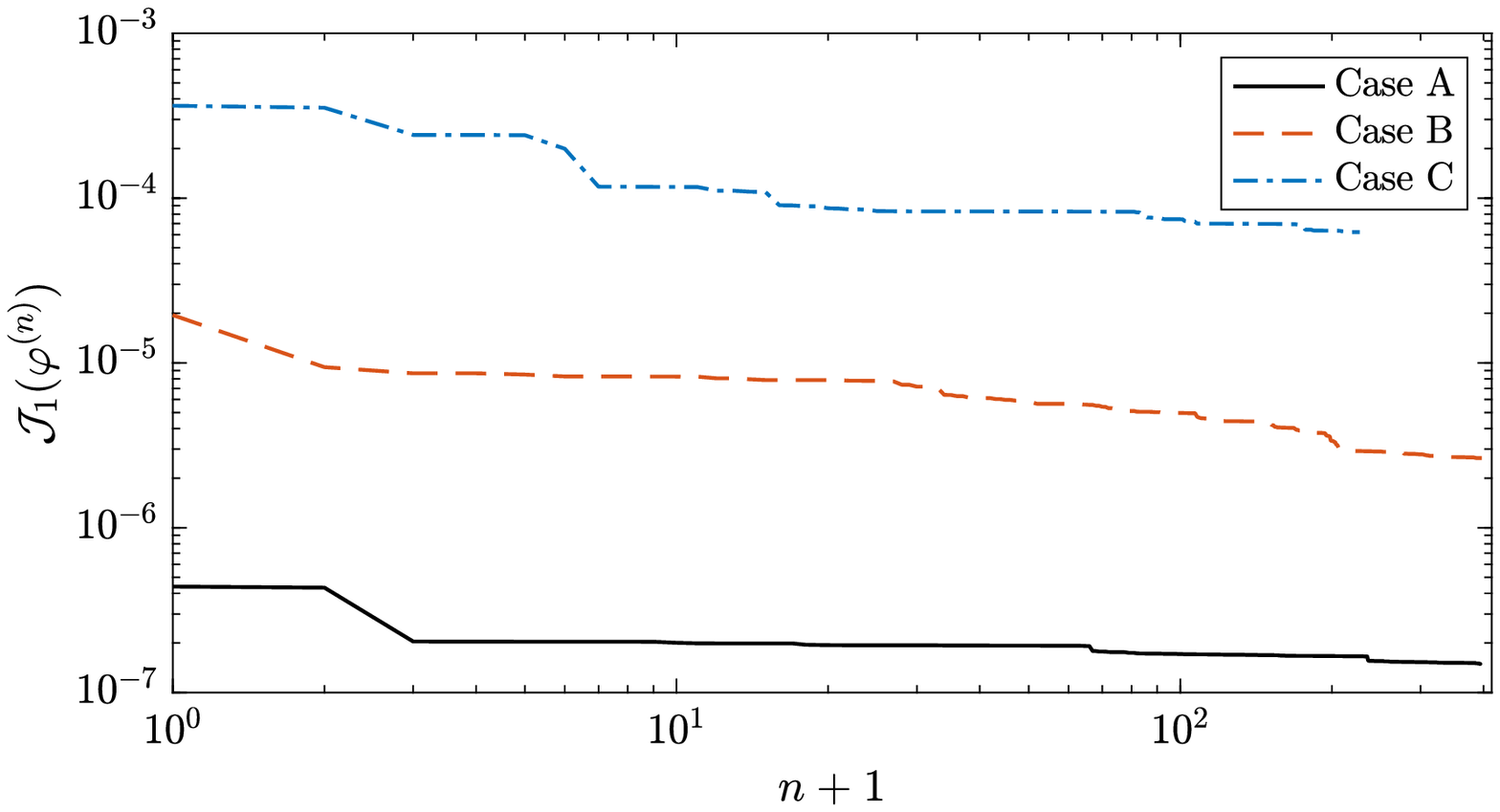}
    \label{fig:costfun_kmax}
  }\quad
  \subfigure[]
  {
    \includegraphics[scale=0.6]{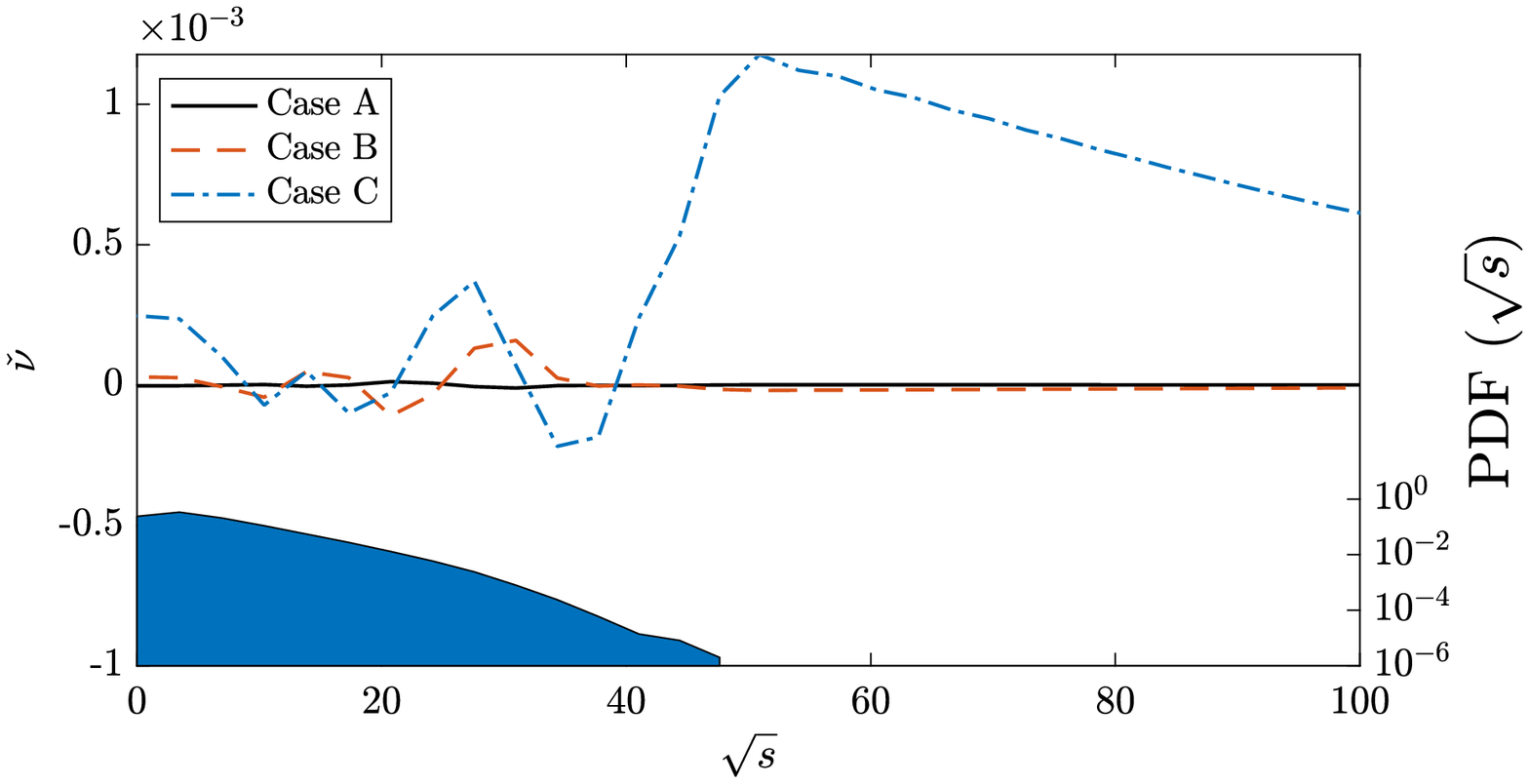}
    \label{fig:OptVisc_kmax}
  }
  \caption{(a) Dependence of the functional
    ${\J_1}({\varphi}^{(n)})$ on the iteration $n$ and (b)
    dependence of the {corresponding} optimal eddy viscosity
    $\nuc$ on $\sqrt{s}$ for cases A, B and C, cf.~Table
    \ref{table:1}.  Panel (b) also shows the PDF of $\sqrt{s}$ in case
    C.}
  \label{fig:ABC}
\end{figure}

Our first set of results addresses the effect of the cutoff wavenumber
$k_c$. They are obtained by solving problem {\cref{eq:minJ} with
  ${j}=1$} for decreasing values of $k_c = 30, 25, 20$ while retaining
fixed values of the regularization parameters $\ell_1,\ell_2$ and a
fixed resolution $N_s$ in the state space $\I$, cf.~cases A, B and C
in Table \ref{table:1}. In each case the optimization problem is
solved using the initial guess {$\varphi_0(s/\smax) \equiv 0$}
corresponding to no closure model at all. The dependence of the error
functional ${\J_1}(\nu^{(n)})$ on iterations $n$ in the three
cases is shown in Figure \ref{fig:ABC}a, where we see that the
mean-square errors between the DNS and the optimal LES increase as the
cutoff wavenumber $k_c$ is decreased {and} the largest relative
reduction of the error is achieved in case C with the smallest
$k_c$. {While minimization in problem \eqref{eq:minJ} is
  performed with respect to the nondimensional function $\varphi$,
  cf.~\eqref{eq:nu}, we focus here on the corresponding optimal eddy
  viscosities $\nuc = \nuc(s)$} shown in Figure \ref{fig:ABC}b.  Since
small values of $s$ are attained more frequently in the flow, cf.~the
probability density function (PDF) of $\sqrt{s}$ embedded in the
figure, the horizontal axis is scaled as $\sqrt{s}$ which magnifies
the region of small values of $s$. We see that for the largest cutoff
wavenumber $k_c = 30$ the optimal eddy viscosity is close to zero over
the entire range of $s$. However, for decreasing $k_c$ the optimal
eddy viscosity exhibits oscillations of increasing magnitude. We note
that values of $s \gtrapprox 50$ occur very rarely in the flow and
hence the gradient \eqref{eq:gradL2} provides little sensitivity
information for $s$ in this range. Thus, the behavior of $\nuc(s)$ for
$s \gtrapprox 50$ is an artifact of the regularization procedure
defined in \eqref{eq:gradH2BVP} and is not physically relevant.

\begin{table}
	\begin{center}
		\begin{tabular}{ |c|c|c|c|c|c|c|c|c|c|c|} 
			\hline
			\rowcolor{Gray}
			 \textbf{Case} & $k_c$ &$N_s$& $\l_1$ & $\ll$&  ${\varphi}_0$ & ${\J_1}({\varphi}_0)$ & ${\J_1}({\varphi}^{(\infty)})$ & $r$  \\ 
			\hline
			\textbf{A} & 30 & $64$ & $10^4$ & $10^3$ & No Closure & {$ 4.398 \times 10^{-7}$} & {$ 1.492 \times 10^{-7}$} & {$8.999 \times 10^{-8}$}   \\ 
			\hline
			\textbf{B} & 25 &$64$ & $10^4$ & $10^3$ & No Closure & {$ 1.951 \times 10^{-5}$} & {$ 2.450 \times 10^{-6}$} & {$1.572 \times 10^{-6}$}  \\ 
			\hline
			\textbf{C} & 20 &$64$ & $10^4$ & $10^3$ & No Closure & {$ 3.635 \times 10^{-4}$} & {$ 6.217 \times 10^{-5}$} & {$4.468 \times 10^{-5}$} \\ 
			\hline
			\textbf{D} & 20 &$128$ & $10^3$ & $10^2$ & Case C & {$ 6.217 \times 10^{-5}$} & {$ 2.001 \times 10^{-5}$} & {$1.239 \times 10^{-5}$}  \\ 
			\hline
			\textbf{E} & 20 & $256$ & $10^1$ & $10^0$ & Case D & {$ 2.333 \times 10^{-5}$} & {$ 1.450 \times 10^{-5}$} & {$8.723 \times 10^{-6}$}  \\ 
			\hline
		\end{tabular}
		\caption{Summary information about the different cases
                  considered {when solving optimization problem
                    \eqref{eq:minJ} with $j=1$.}}
		\label{table:1}
	\end{center}
\end{table}

\begin{figure}\centering
  \subfigure[]
  {
    \includegraphics[scale=0.6]{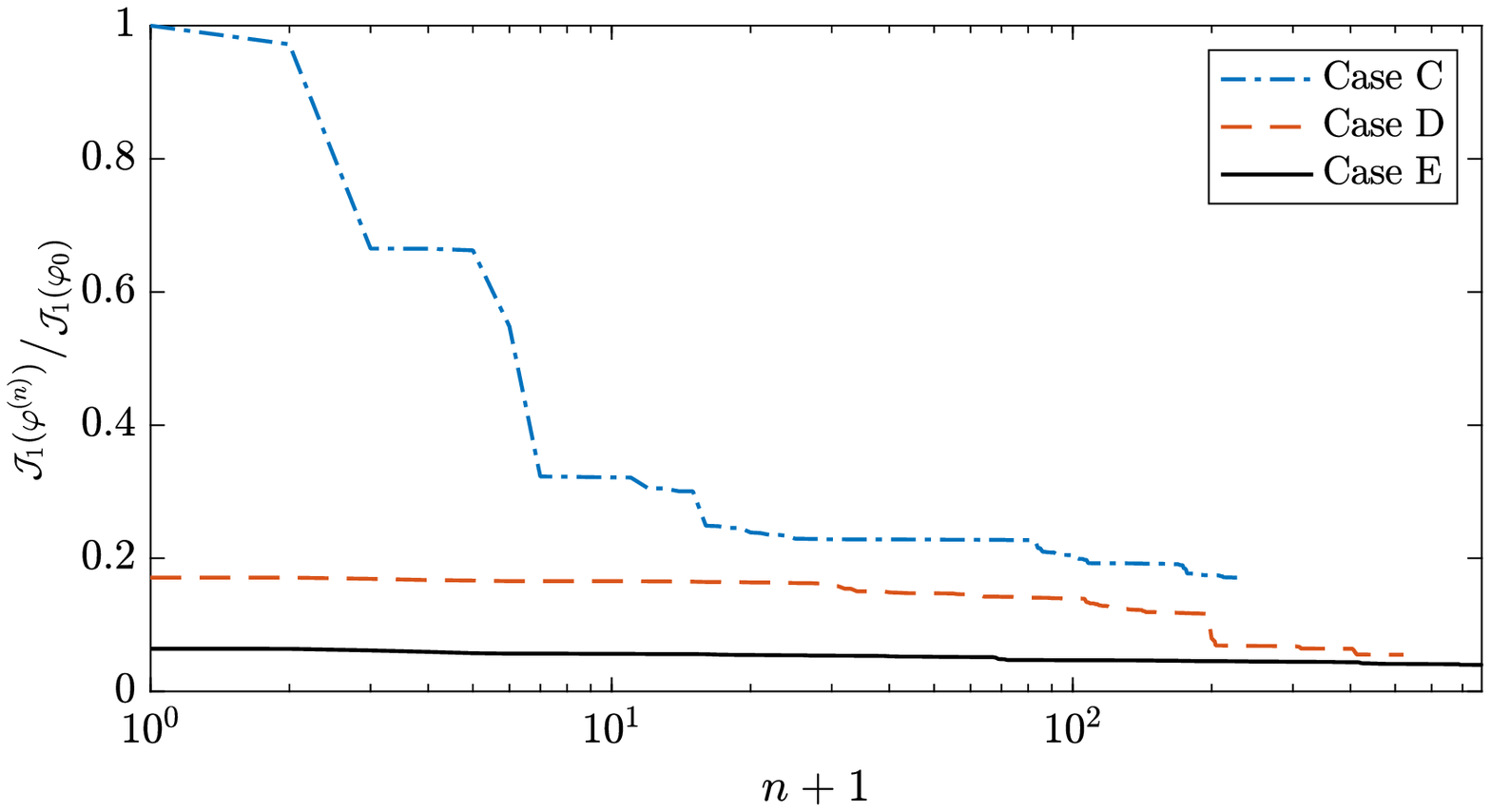}
    \label{fig:costfun}
  }\quad
  \subfigure[]
  {
    \includegraphics[scale=0.6]{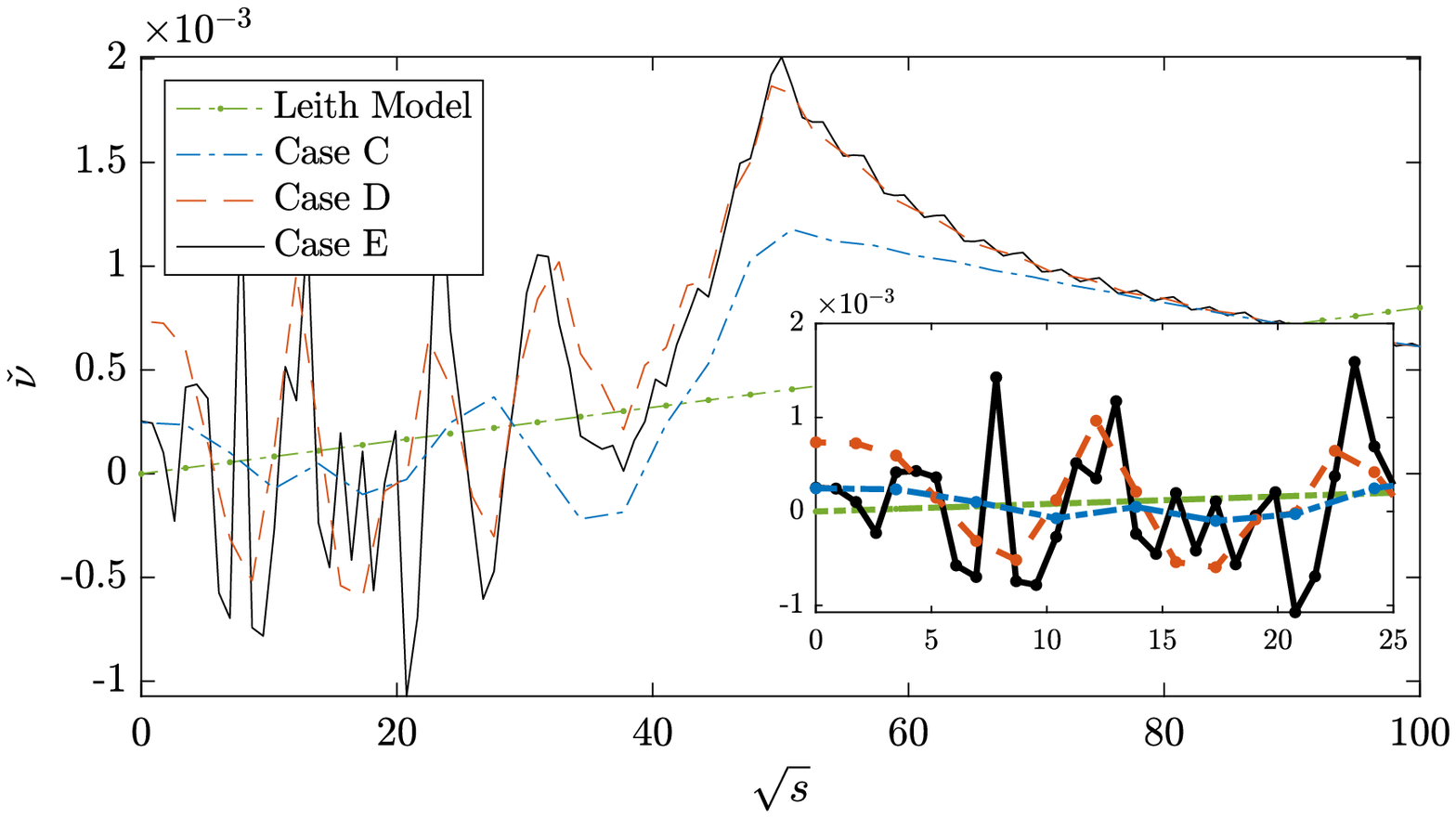}
    \label{fig:OptVisc}
  }
  \caption{(a) Dependence of the normalized functional
    ${\J_1}({\varphi}^{(n)})/{\J_1}({\varphi}_0)$,
    with ${\J_1}({\varphi}_0)$ from case C, on the iteration
    $n$ and (b) dependence of the {corresponding} optimal eddy
    viscosity $\nuc$ on $\sqrt{s}$ for cases C, D and E, cf.~Table
    \ref{table:1}. The inset in panel (b) shows magnification of the
    region $\sqrt{s} \in [0,25]$. Panel (b) also shows the Leith model
    with $k_c =20$ and the eddy viscosity {$\nu_L(s)$,
      cf.~\eqref{eq:nu0}.}}
  \label{fig:CDE}
\end{figure}

In order to provide additional insights about the properties of the
optimal eddy viscosity, our second set of results is obtained as
solutions of problem {\cref{eq:minJ} with ${j}=1$ using} a fixed
$k_c = 20$ and progressively reduced regularization achieved by
decreasing the parameters $\ell_1,\ell_2$ while simultaneously
refining the resolution $N_s$ in the state space $\I$, cf.~cases C, D
and E in Table \ref{table:1}. Optimization problems with weaker
regularization are solved using the optimal eddy viscosity obtained
with stronger regularization as the initial guess. From the normalized
error functionals shown as functions of iterations in Figure
\ref{fig:CDE}a, we see that as regularization is reduced, the
mean-square errors between the optimal LES and the DNS become smaller
and approach a certain nonzero limit, cf.~Table \ref{table:1}. As is
evident from Figure \ref{fig:CDE}b, this is achieved with the
corresponding optimal eddy viscosities developing oscillations with an
ever increasing frequency.  More precisely, each time the
regularization parameters $\ell_1,\ell_2$ are reduced and the
resolution $N_s$ is refined, a new oscillation with a higher frequency
appears in the optimal eddy viscosity $\nuc(s)$ (in fact, in each
case, this is the highest-frequency oscillation which can be
represented on a grid with $N_s$ points).

\begin{figure}\centering
  \subfigure[]
  {
    \includegraphics[scale=0.6]{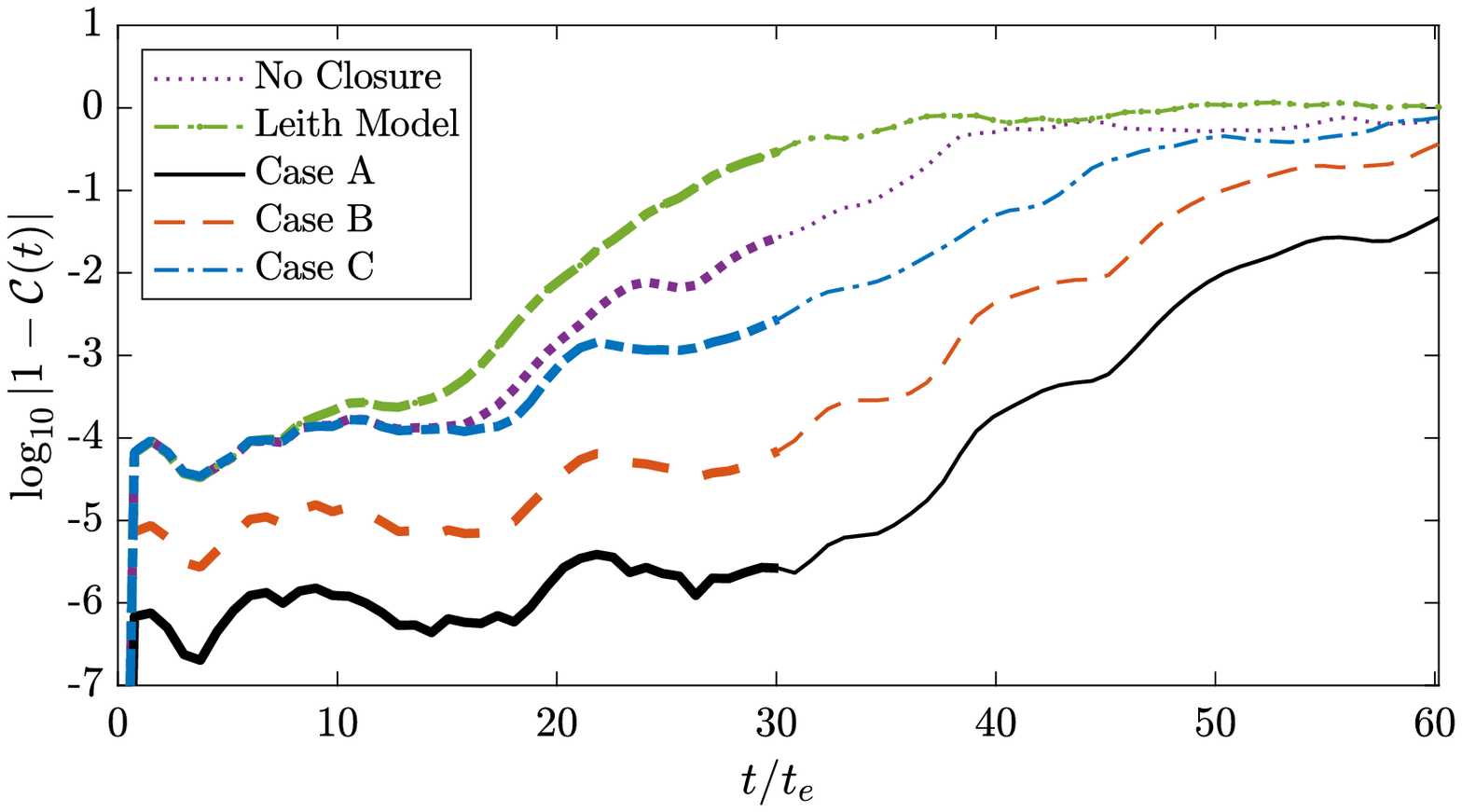}
    \label{fig:corr_kmax}
  }\quad
  \subfigure[]
  {
    \includegraphics[scale=0.6]{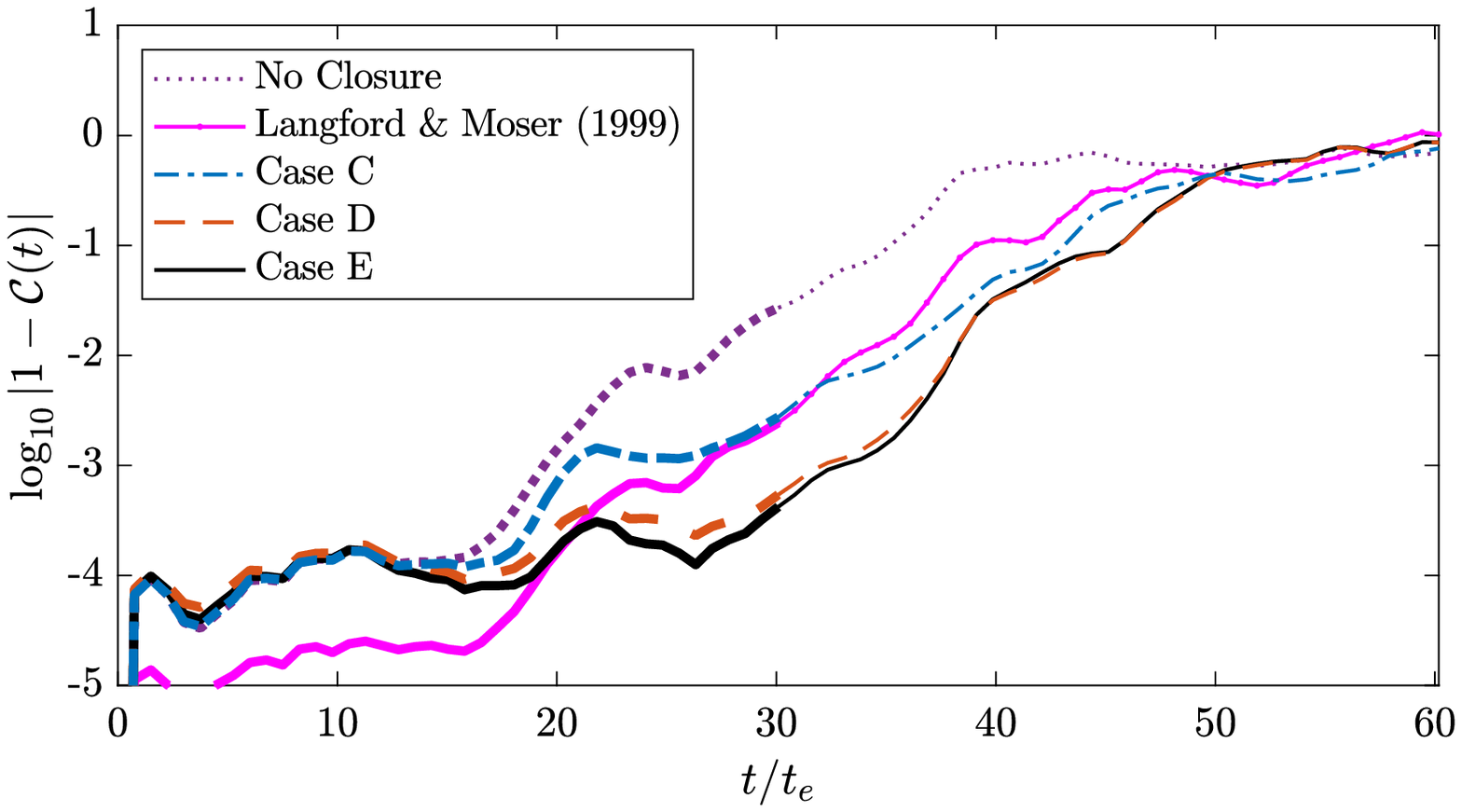}
    \label{fig:corr}
  }
  \caption{Adjusted normalized correlations \cref{eq:C} for the LES
    with (a) no closure and the optimal eddy viscosity in cases A, B
    and C, and (b) no closure and the optimal eddy viscosity in cases
    C, D and E. The correlation is also shown for the Leith model with
    {$k_c =20$ and} the eddy viscosity {$\nu_L(s)$,
      cf.~\eqref{eq:nu0},} in (a) and for an optimal closure model
    based on the stochastic estimator \citep{langford1999optimal} in
    (b).  Thick and thin lines correspond to, respectively, time in
    the ``training window'' ($t \in [0, T]$) and beyond this window
    ($t \in (T, 2T]$).}
  \label{fig:C}
\end{figure}

In order to assess how well the solutions of the LES system
\eqref{eq:LES} with the optimal eddy viscosities $\nuc$ shown in
Figures \ref{fig:ABC}b and \ref{fig:CDE}b approximate the solution of
the Navier-Stokes system \eqref{eq:2DNS}, in Figures \ref{fig:C}a and
\ref{fig:C}b we show the time evolution of the quantity
$\log_{10}|1-\C(t)|$ where
\begin{equation}
  \C(t) := \frac{1}{||\widetilde{w}(t)||_{L^2(\Omega)} \, ||\tw(t)||_{L^2(\Omega)} } \int_{\Omega} \widetilde{w}(t, \x) \, \tw(t, \x) \ d\Omega \label{eq:C}
\end{equation}
is the normalized correlation between the two flows. For a more
comprehensive assessment, these results are shown for $t \in [0,2T]$,
i.e., for times up to twice longer than the ``training window''
$[0,T]$ used in the optimization problem \cref{eq:minJ}.  In Figure
\ref{fig:C}b we also present the results obtained {for $k_c=20$} with
an optimal closure model based on the linear stochastic estimator
introduced by \citet{langford1999optimal}. Since at early times
correlation $\C(t)$ reveals exponential decay corresponding to the
exponential divergence of the LES flow from the DNS, this effect can
be quantified by approximating the correlation as
$\C(t) \approx \bar{\C}(t) := \C_0 e^{- r t}$, where {$\C_0 = 1$
  follows the fact that $\tw_0 \equiv \widetilde{w}_0$}, whereas the
decay rate $r$ is obtained from a least-squares fit over the time
window $[0,T]$. The decay rates $r$ obtained in this way are collected
in Table \ref{table:1}.

\begin{figure}\centering
  \subfigure[]
  {
    \includegraphics[scale=0.48]{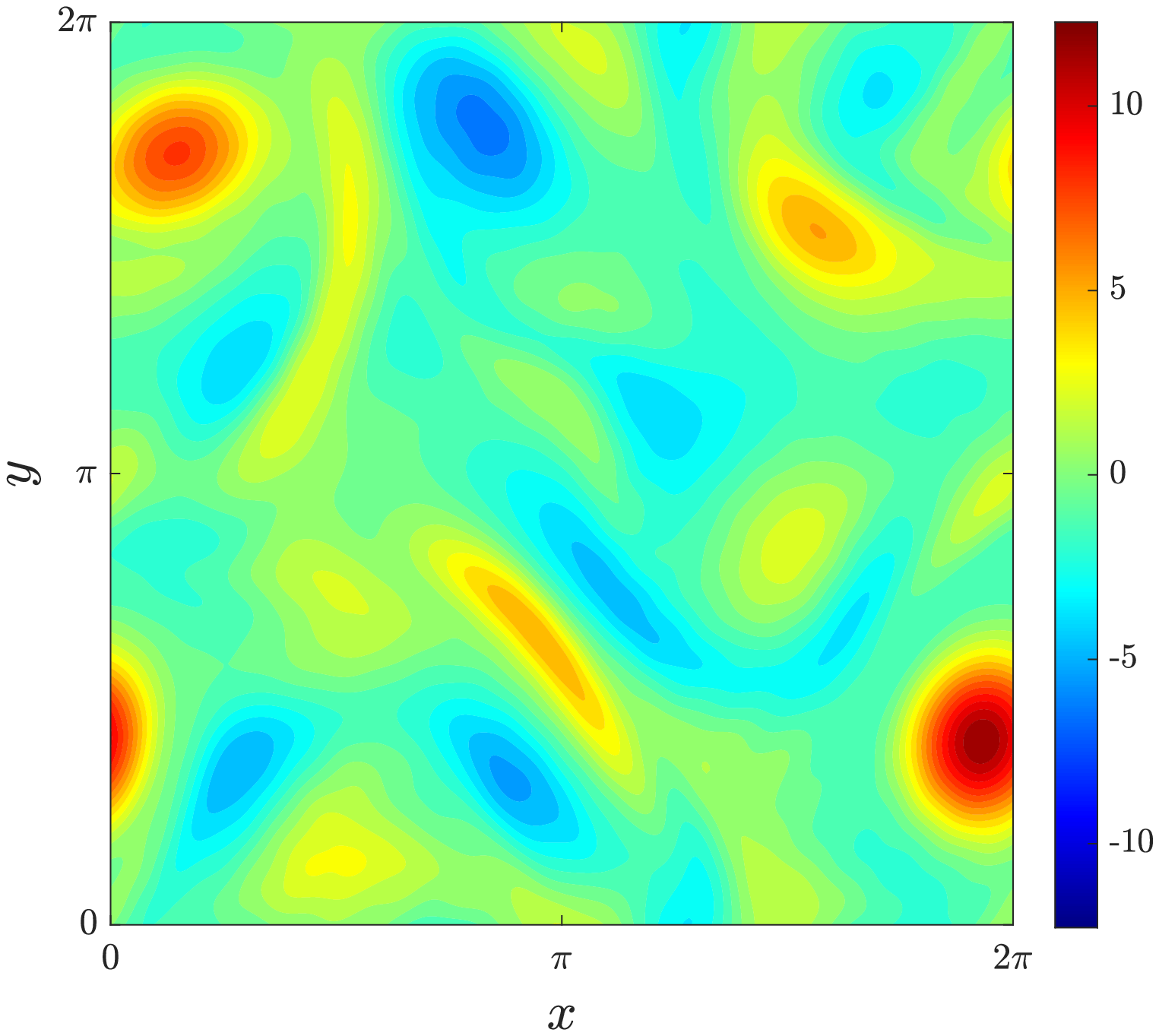}
  }\quad
  \subfigure[]
  {
    \includegraphics[scale=0.48]{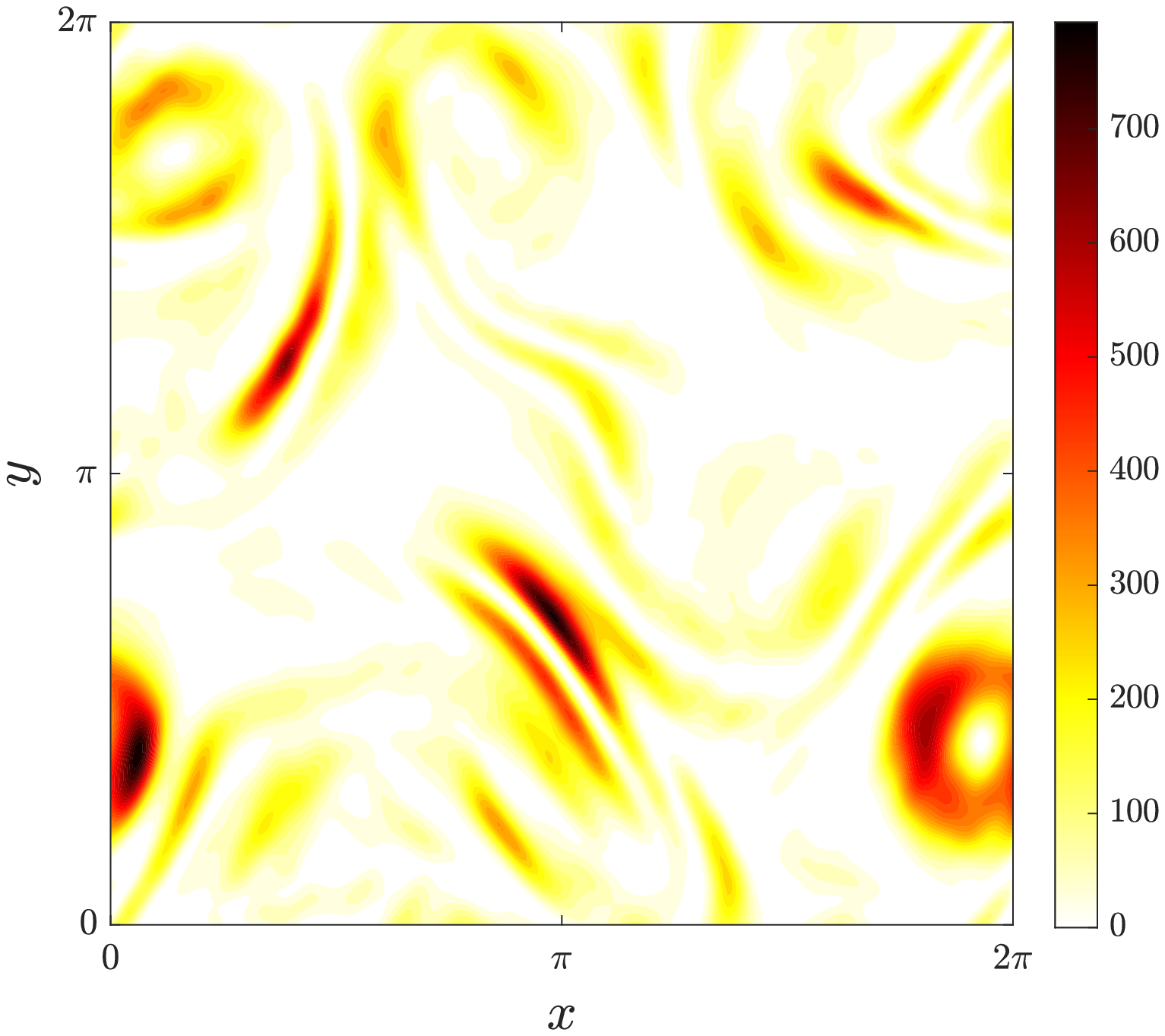}
  }\quad
  \subfigure[]
  {
    \includegraphics[scale=0.48]{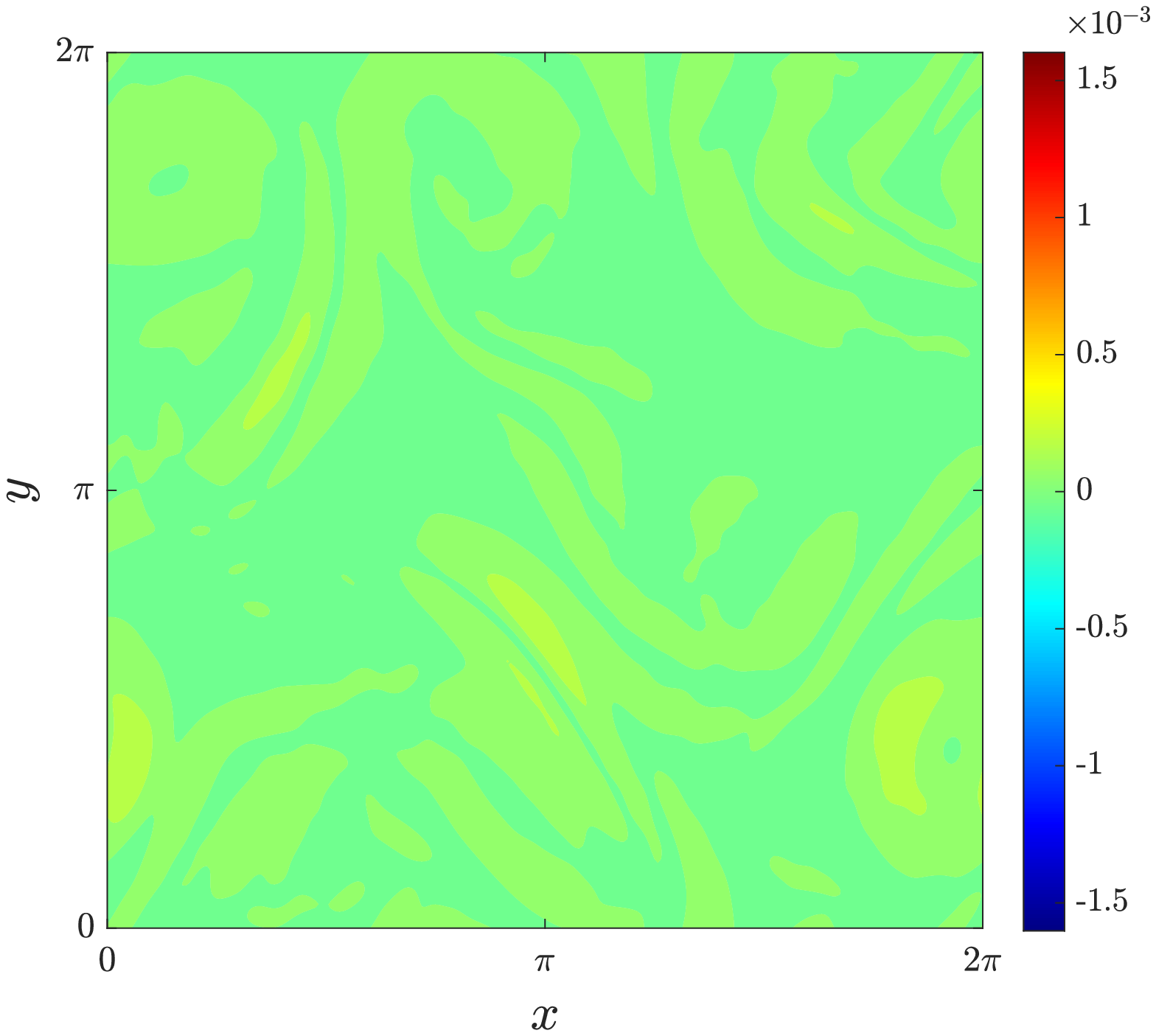}
  }\quad
  \subfigure[]
  {
    \includegraphics[scale=0.48]{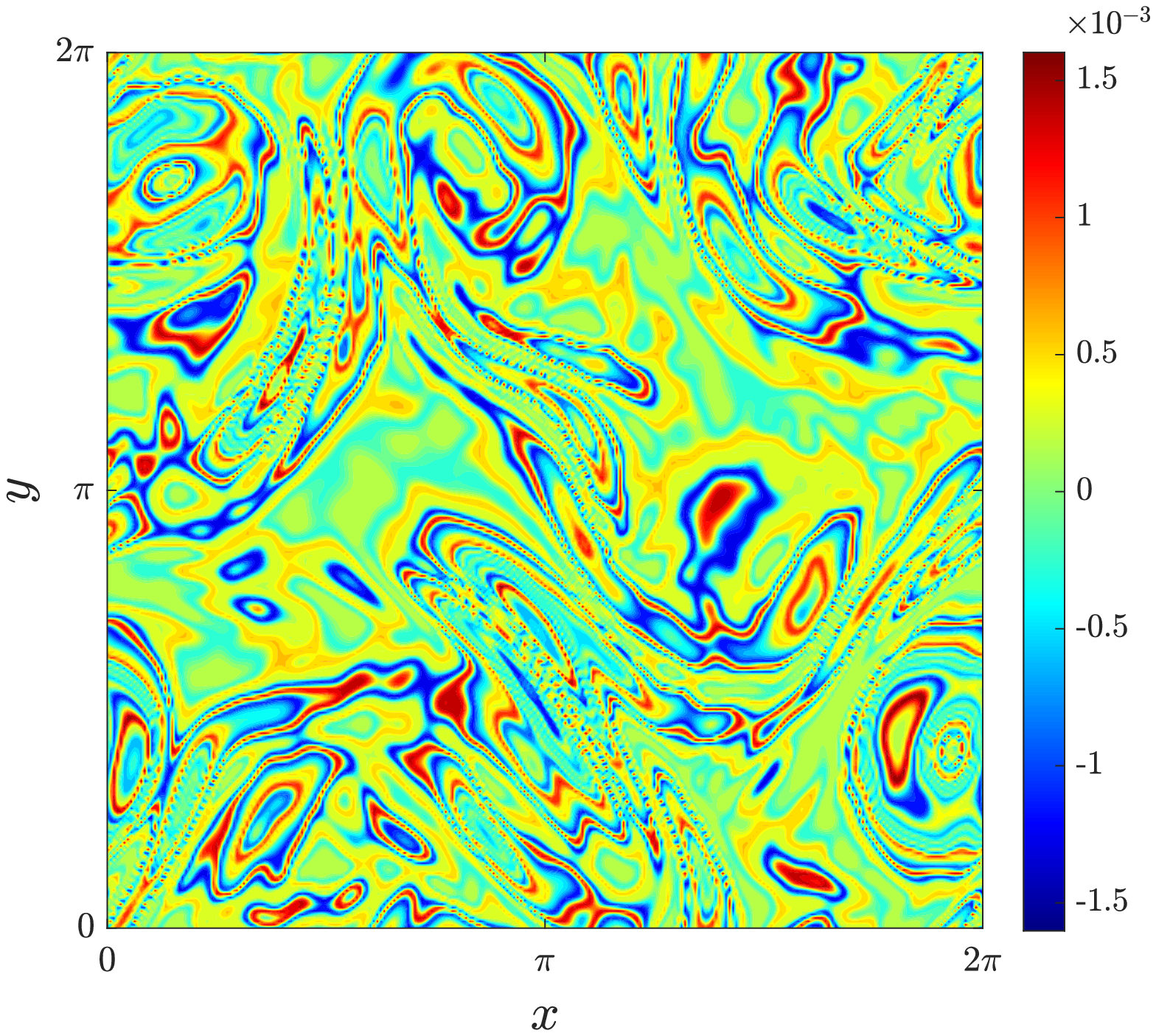}
  }
  \caption{For case E we show: (a) the vorticity field $\tw(T,\x)$,
    $\x \in \Omega$, (b) the corresponding state variable $s(T,\x)$,
    cf.~\eqref{eq:nu}, and the spatial distribution of (c) the
    {eddy viscosity $\nu_L(s(T,\x))$ in the Leith model,
      cf.~\eqref{eq:nu0} with $\delta = 0.02$,} and (d) the optimal
    eddy viscosity $\nuc(s(T,\x))$, cf.~Figure \ref{fig:CDE}b, all
    shown at the end of the training window for $t = T$. For better
    comparison the same color scale is used in panels (c) and (d).}
  \label{fig:nuc}
\end{figure}

Finally, in order to provide insights about how the closure model with
the optimal eddy viscosity acts in the physical space, in Figures
\ref{fig:nuc}a, \ref{fig:nuc}b and \ref{fig:nuc}d we show the
vorticity field $\tw(T,\x)$, the corresponding state variable
$s(T,\x)$ cf.~\eqref{eq:nu}, and the spatial distribution
$\nuc(s(T,\x))$ of the optimal eddy viscosity obtained in case E; {for
  comparison, the spatial distribution of the {eddy viscosity
    $\nu_L(s(T,\x)) $ in the Leith model, cf.~\eqref{eq:nu0} with
    $\delta = 0.02$}, is shown in Figure \ref{fig:nuc}c} (the fields
are shown in the entire domain, i.e., for $\x \in \Omega$, at the end
of the training window). We see that while the vorticity and
state-variable fields vary smoothly, this is also the case for the
spatial distribution of the {eddy viscosity $\nu_L(s(T,\x))$ in the
  Leith model}. On the other hand, the spatial distribution of the
optimal eddy viscosity {$\nuc(s(T,\x))$} exhibits rapid variations,
which is consistent with the results presented in Figure
\ref{fig:CDE}b. In particular, positive and negative values of
$\nuc(s(T,\x))$, corresponding to localized dissipation and injection
of enstrophy, tend to form concentric bands in some low-vorticity
regions of the flow domain.  The time evolution of the vorticity field
in the DNS, LES with no closure model and LES with the optimal eddy
viscosity (case E) are available together with an animated version of
Figure \ref{fig:C}b as a movie
\href{https://youtu.be/Xhztbo-VglM}{on-line}.  An animation
representing the time evolution of the fields shown in Figure
\ref{fig:nuc} for $t \in [0,{2}T]$ is available as a movie
\href{https://youtu.be/qXHWebW1hsM}{on-line}.

\FloatBarrier

{\subsection{Matching the DNS in an Average Sense ---
    Results for the Optimization Problem with Error Functional  \eqref{eq:J2} \label{sec:resultsJ2}}}

{ Now we review the results obtained by solving optimization
  problem \eqref{eq:minJ} for ${j}=2$ with a fixed cutoff wavenumber
  $k_c = 20$ and with two sets of parameters determining
  regularization ($\ell_1$ and $\ell_2$) and the resolution in the state
  space $\I$ ($N_s$), cf.~cases F and G in Table \ref{table:2}. We
  remark that the regularization performed in the {present} problem is
  less aggressive than in the problem discussed in Section
  \ref{sec:resultsJ1}.  As shown in \Cref{fig:FG}a, the normalized
  error functional converges to a local minimum in only a few
  iterations and, as the regularization is reduced, a larger reduction
  of the error functional is obtained.  However, as is evident from
  \Cref{fig:FG}b, this is achieved with optimal eddy viscosities much
  better behaved than the optimal eddy viscosities found by solving
  the optimization problem discussed in Section \ref{sec:resultsJ1},
  even though a weaker regularization is now applied, cf.~Table
  \ref{table:2} (the obtained optimal eddy viscosity exhibits more
  small-scale variability in case G than in case F, but the difference
  is not significant).

  The difference between the time-averaged vorticity spectra
  \eqref{eq:E} is the LES with no closure, LES with the optimal
  closure $\nuc$ (cases F and G) and in the filtered DNS is shown in
  Figure \ref{fig:Vort_J} as a function of the wavenumber $k$ (this
  quantity is related to the integrand expression in the error
  functional \eqref{eq:J2}). We see that when the optimal eddy
  viscosity $\nuc$ is used in the LES, this error is reduced,
  especially at low wavenumbers $k$. On the other hand, the evolution
  of the quantity $\log_{10}|1-\C(t)|$, cf.~\eqref{eq:C}, shown for
  the same cases in \Cref{fig:CVort} demonstrates that, in contrast to
  Figure \ref{fig:C}, in the present problem the LES flows equipped
  with the optimal eddy viscosity do not achieve a better
  pointwise-in-space accuracy with respect to the DNS than the LES
  flow with no closure model.

  Finally, we show the vorticity field $\tw(T, \x)$, the corresponding
  state variable $s(T, \x)$, cf.~\eqref{eq:nu}, the spatial
  distribution $\nuc(s(T,\x))$ of the optimal eddy viscosity obtained
  in case G, and for comparison, the spatial distribution of the eddy
  viscosity $\nu_L(s(T,\x)) $ in the Leith model, cf.~\eqref{eq:nu0},
  in Figures \ref{fig:nucVort}a, \ref{fig:nucVort}b,
  \ref{fig:nucVort}d, and \ref{fig:nucVort}c, respectively. We remark
  that the spatial distribution of the optimal eddy viscosity in
  Figure \ref{fig:nucVort}d is now significantly smoother than the
  distribution of the optimal eddy viscosity obtained in the first
  formulation by solving optimization problem \eqref{eq:minJ} with
  ${j}=1$, cf.~Figure \ref{fig:nuc}d. An animated version of Figure
  \ref{fig:nucVort} illustrating the evolution of the fields for $t
  \in [0,{2}T]$ {is} available as a movie
\href{https://youtu.be/Yr3I_aPVIRs}{on-line}.

\begin{table}
	\begin{center}
		\begin{tabular}{ |c|c|c|c|c|c|c|c|c|c|c|} 
			\hline
			\rowcolor{Gray}
			 {\textbf{Case}} & $k_c$ &$N_s$& $\l_1$ & $\ll$&  $\varphi_0$ & $\J_2(\varphi_0)$ & $\J_2(\varphi^{(\infty)})$ & $r$  \\ 
			\hline
			{\textbf{F}} & 20 & $256$ & $10^1$ & $10^0$ & No Closure & {$ 6.736 \times 10^{-2}$} & {$ 8.876 \times 10^{-3}$} & {$2.882 \times 10^{-4}$}  \\ 
			\hline
			{\textbf{G}} & 20 &$512$ & $10^{-1}$ & $10^{-2}$ & No Closure & {$ 6.736 \times 10^{-2}$} & {$ 6.286 \times 10^{-3}$} & {$1.685 \times 10^{-4}$}  \\ 
			\hline
		\end{tabular}
		\caption{{Summary information about the different
                    cases considered when solving optimization problem
                    \eqref{eq:minJ} with ${j}=2$.}}
		\label{table:2}
	\end{center}
\end{table}

\begin{figure}\centering
  \subfigure[]
  {
    \includegraphics[scale=0.6]{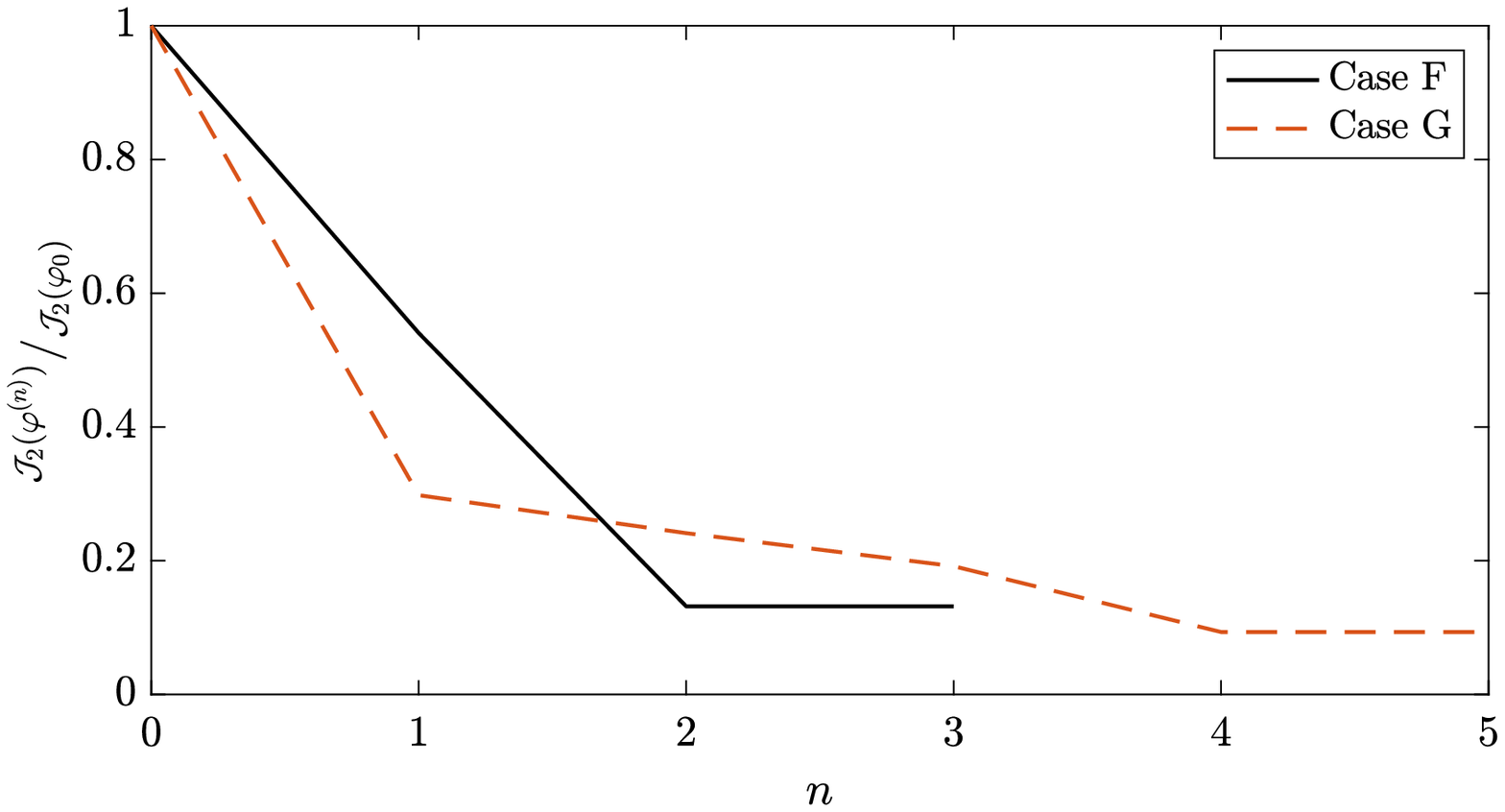}
    \label{fig:costfun2}
  }\quad
  \subfigure[]
  {
    \includegraphics[scale=0.6]{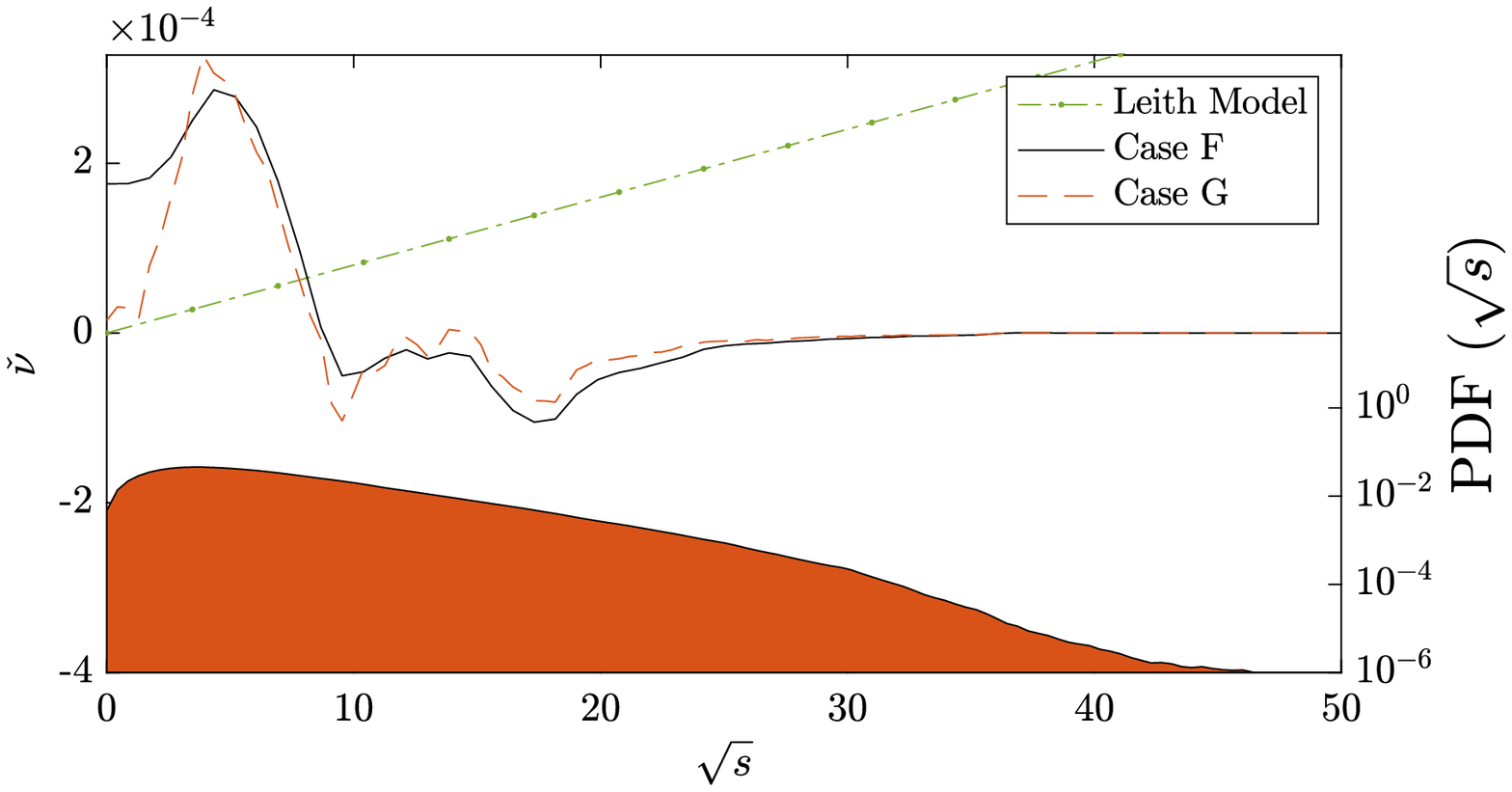}
    \label{fig:OptVisc2}
  }
  \caption{{(a) Dependence of the normalized functional
      $\J_2({\varphi}^{(n)}) /\J_2(\varphi_0)$ on the iteration $n$
      and (b) dependence of the {corresponding} optimal eddy viscosity
      $\nuc$ on $\sqrt{s}$ for cases F and G, cf.~Table \ref{table:2}.
      Panel (b) also shows the PDF of $\sqrt{s}$ in case G.}}
  \label{fig:FG}
\end{figure}

\begin{figure}\centering
    \includegraphics[scale=0.6]{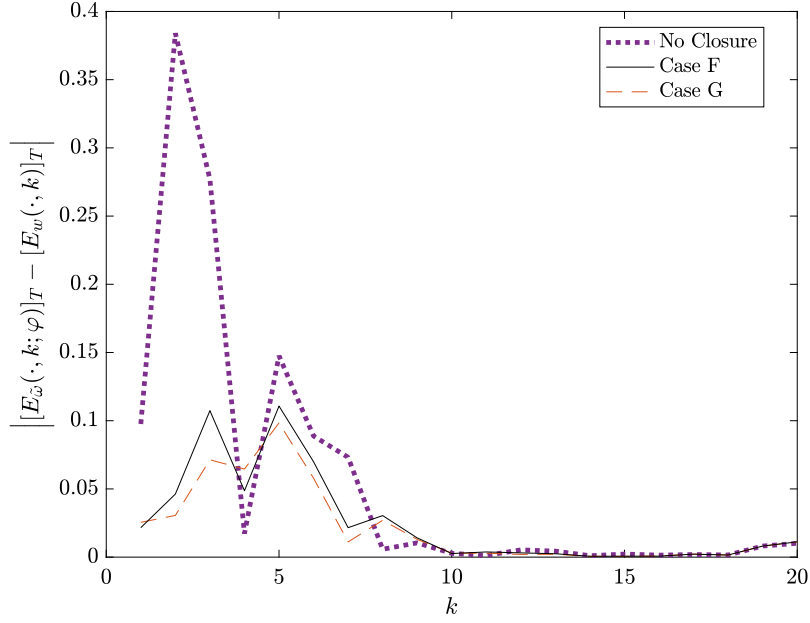}
    \caption{{The difference between time-averaged vorticity
        spectra \eqref{eq:E} in the filtered DNS and in the LES with
        no closure and with the optimal eddy viscosity $\nuc$ obtained
        in cases F and G, cf.~Table \ref{table:2}, as function of the
        wavenumber $k$.}}
  \label{fig:Vort_J}
\end{figure}

\begin{figure}\centering
    \includegraphics[scale=0.6]{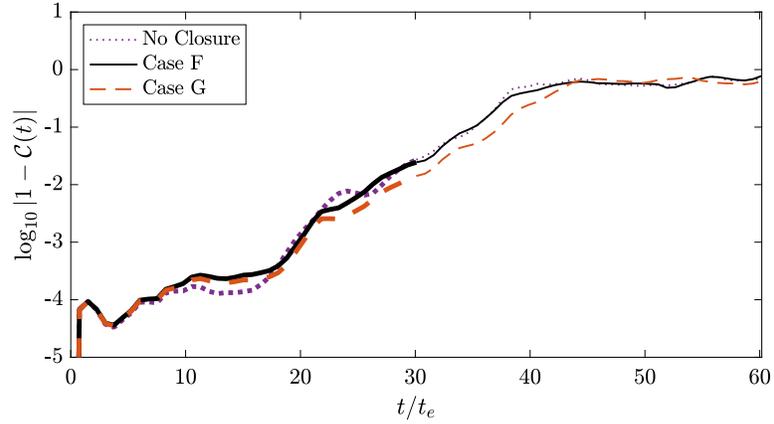}
    \caption{{Adjusted normalized correlations \cref{eq:C} for
        the LES with no closure and the optimal eddy viscosity in
        cases F and G. Thick and thin lines correspond to,
        respectively, time in the ``training window'' ($t \in [0, T]$)
        and beyond this window ($t \in (T, 2T]$).}}
  \label{fig:CVort}
\end{figure}

\begin{figure}\centering
  \subfigure[]
  {
    \includegraphics[scale=0.48]{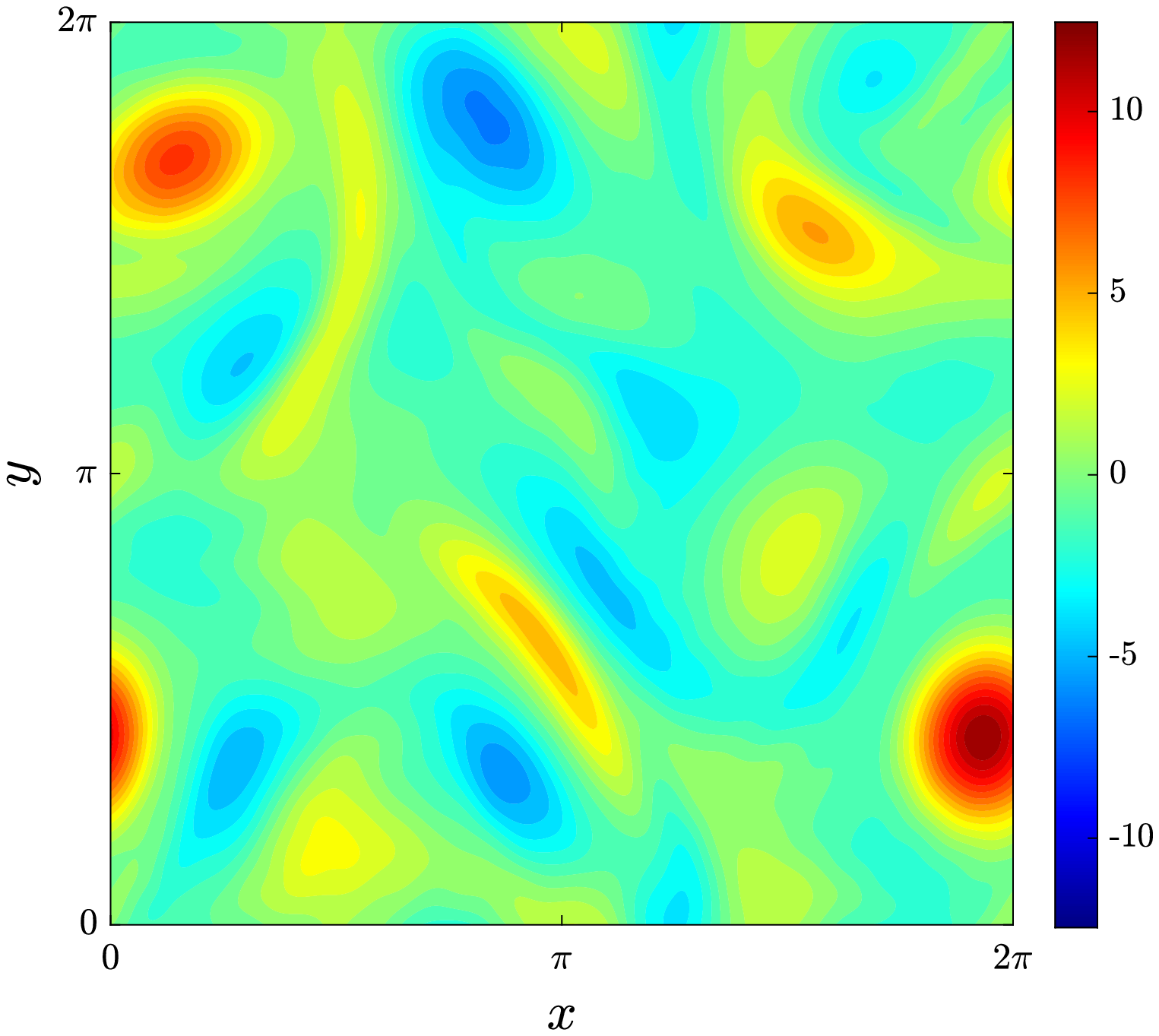}
  }\quad
  \subfigure[]
  {
    \includegraphics[scale=0.48]{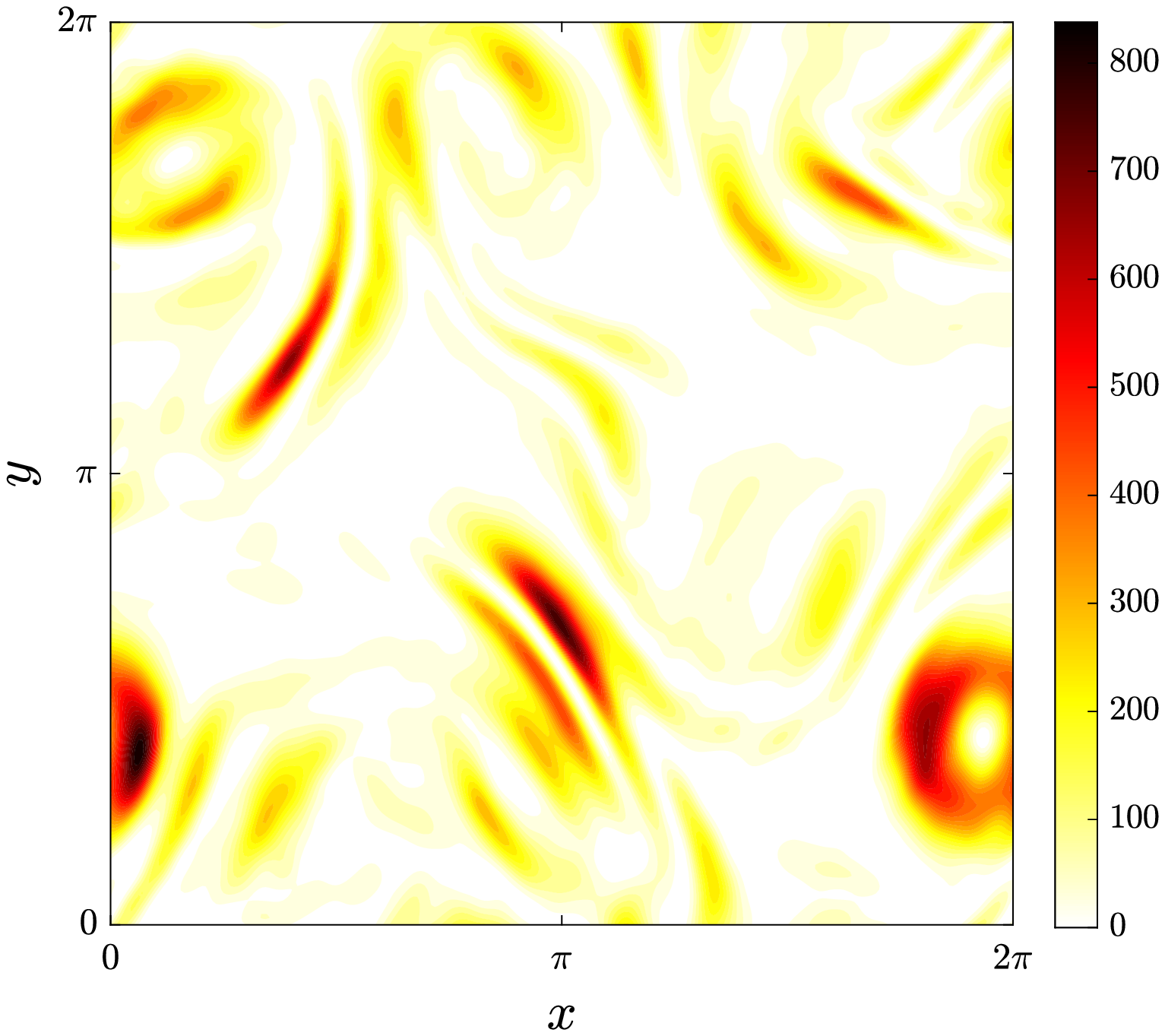}
  }\quad
  \subfigure[]
  {
    \includegraphics[scale=0.48]{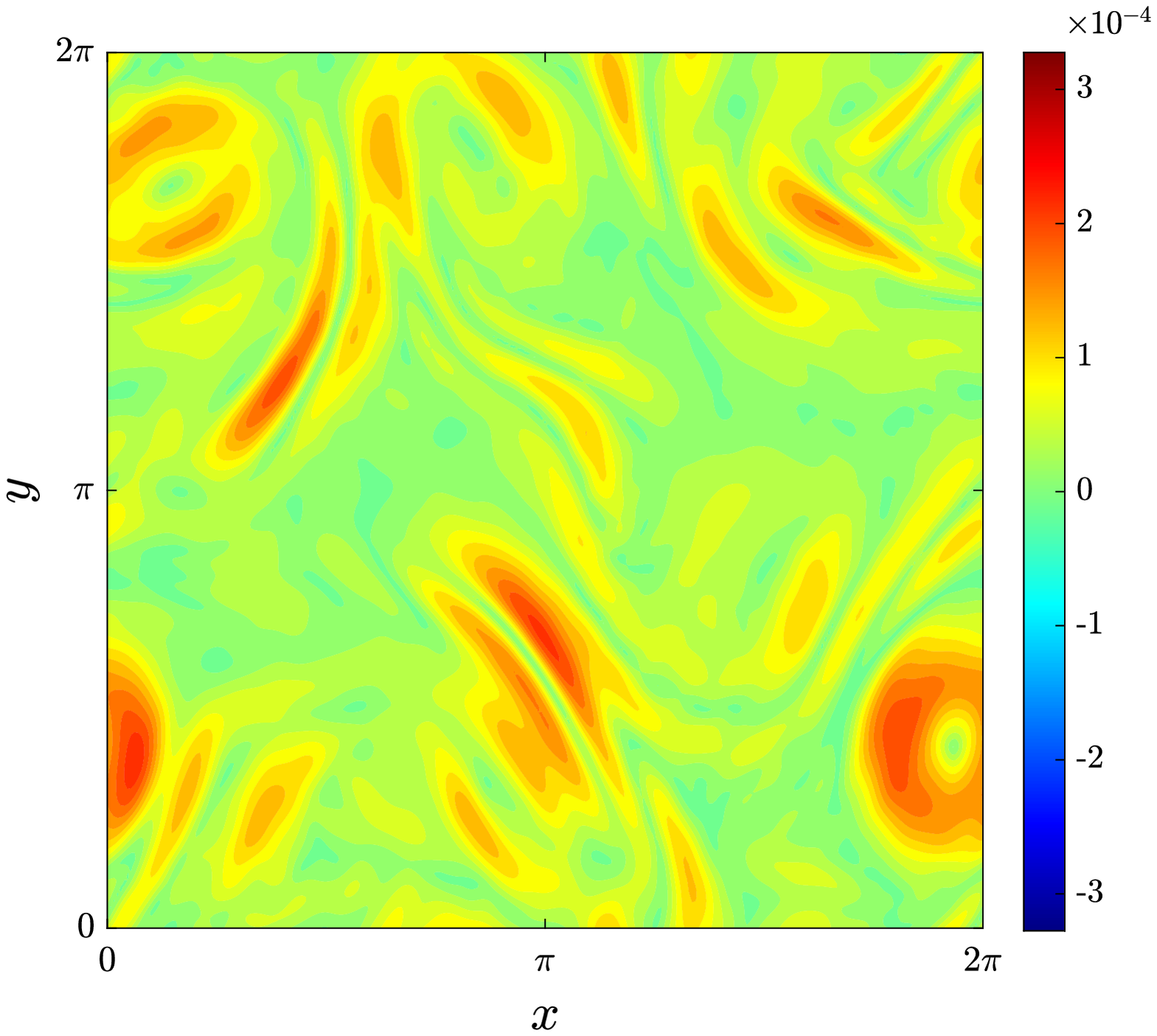}
  }\quad
  \subfigure[]
  {
    \includegraphics[scale=0.48]{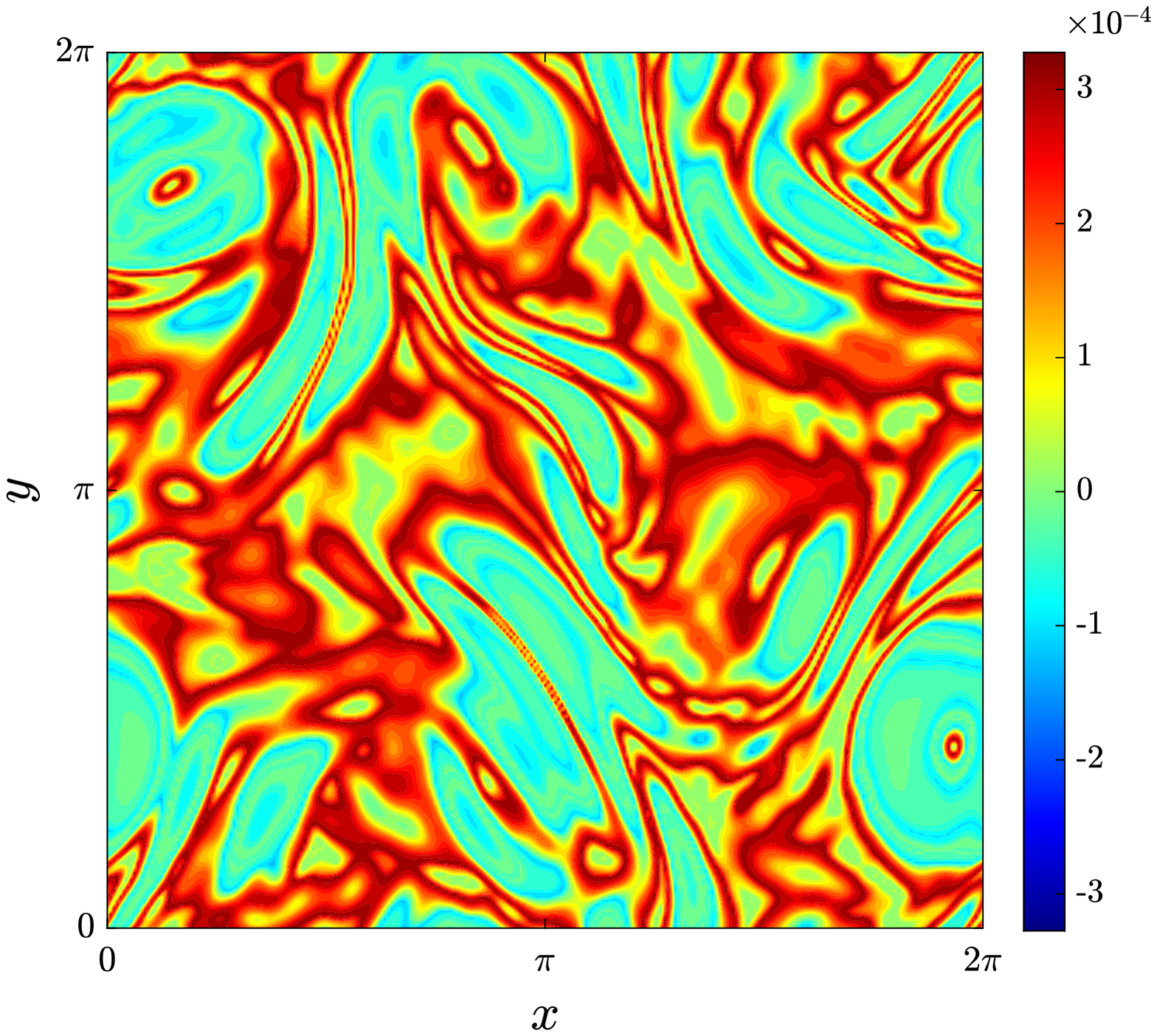}
  }
  \caption{{For case G we show: (a) the vorticity field
      $\tw(T,\x)$, $\x \in \Omega$, (b) the corresponding state
      variable $s(T,\x)$, cf.~\eqref{eq:nu}, and the spatial
      distribution of (c) the eddy viscosity $\nu_L(s(T,\x))$ in the
      Leith model, cf.~\eqref{eq:nu0} with $\delta = 0.02$, and (d)
      the optimal eddy viscosity $\nuc(s(T,\x))$, cf.~Figure
      \ref{fig:FG}b, all shown at the end of the training window for
      $t = T$. For better comparison the same color scale is used in
      panels (c) and (d).}}
  \label{fig:nucVort}
\end{figure}

\clearpage
\section{Discussion and Conclusions \label{sec:final}}

{In this study we have considered the question of fundamental
  limitations on the performance of eddy-viscosity closure models for
  turbulent flows. We focused on the Leith model for 2D LES for which
  we sought optimal eddy viscosities that subject to minimum
  assumptions would result in the least mean-square error between the
  corresponding LES and the {filtered} DNS. Such eddy viscosities were
  found as minimizers of a PDE-constrained optimization problem with a
  nonstandard structure which was solved using a suitably adapted
  adjoint-based gradient approach \citep{Matharu2020}. A key element
  of this approach was a regularization strategy involving the
  length-scale parameters $\ell_1$ and $\ell_2$ in the Sobolev
  gradients, cf.~\eqref{eq:gradH2BVP}. {The approach proposed is
    admittedly rather technically involved which may limit its
    practical applicability to construct new forms of the eddy
    viscosity, but its value is in making it possible to
    systematically characterize the best possible performance of
    different types of closure models.}

  Our main finding {in \Cref{sec:resultsJ1}} is that with a fixed
  cutoff wavenumber $k_c$ the LES with an optimal eddy viscosity
  $\nuc$ matches the DNS increasingly well as the regularization in
  the solution of the optimization problem is reduced, cf.~Figure
  \ref{fig:CDE}a. This is quantified by a reduction of the rate of
  exponential decay of the correlation between the corresponding LES
  and the DNS, cf.~Figure \ref{fig:C}b and Table \ref{table:1}. This
  optimal performance of the closure model is achieved with eddy
  viscosities $\nuc(s)$ rapidly oscillating with a frequency
  increasing as the regularization parameters are reduced. From this
  we conclude that in the limit of vanishing regularization parameters
  and an infinite numerical resolution the optimal eddy viscosity
  would be undefined as it would exhibit oscillations with an
  unbounded frequency. Thus, from the mathematical point of view, the
  problem of finding an optimal eddy viscosity in the absence of
  regularization is ill-posed. In practical terms, this means that the
  ``best'' eddy viscosity for the Leith model does not exist.

  The optimal performance of the LES is realized by a rapid variation
  of the eddy viscosity $\nuc(s)$ which oscillates between positive
  and negative values as $s$ changes, cf.~Figure \ref{fig:CDE}b,
  resulting in the dissipation and injection of the enstrophy
  occurring in the physical domain in narrow alternating bands,
  cf.~Figure \ref{fig:nuc}d. We note that a somewhat similar behavior
  was also observed in \citep{Maulik2020} where the authors used
  machine learning methods to determine pointwise estimates of eddy
  viscosity which exhibited oscillations between positive and negative
  values. This behavior can be understood in physical terms based on
  relations \eqref{eq:M}--\eqref{eq:Leith} which can be interpreted as
  defining the eddy viscosity in terms of the space- and
  time-dependent DNS field, but the problem is severely
  overdetermined. Thus, some form of relaxation is needed to determine
  $\nu$ and the proposed optimization approach with its inherent
  regularization strategy is one possibility.
    
  In addition, the optimal eddy viscosities {found here} have the
  property that $\nuc(0) > 0$, in contrast to what is typically
  assumed in the Leith model where $\nu(0) = 0$ \citep{Maulik2017}. In
  contrast to the behavior observed in Figure \ref{fig:nuc}d, standard
  eddy viscosity closure models are usually assumed to be strictly
  dissipative \cite{rodi2013large}, which is reflected in the fact
  that the eddy viscosity is non-negative as in Figure \ref{fig:nuc}c.
  {We add that we have also considered finding optimal eddy
    viscosities by matching against the unfiltered DNS field, i.e.,
    using $w(t, \x)$ in the error functional \eqref{eq:J} instead of
    $\widetilde{w(}t, \x)$, however, this approach produced results
    very similar to the ones reported above.}  As is evident from
  Figure \ref{fig:C}b, the performance of the LES with optimal eddy
  viscosities compares favourably to the LES with an optimal closure
  model proposed by \citet{langford1999optimal} based on a stochastic
  estimator, which has a less restrictive structure than the Leith
  model.

  {The optimal eddy viscosities constructed in
    \Cref{sec:resultsJ1} to maximize the pointwise match against the
    filtered DNS are unlikely to be useful in practice due to their
    highly irregular behaviour which is difficult to resolve using
    finite numerical precision.  On the other hand, the second
    formulation studied in \Cref{sec:resultsJ2} where optimal eddy
    viscosities were determined by matching predictions of the LES
    against the time-averaged vorticity spectrum of the DNS for small
    wavenumbers lead to a much better behaved optimization problem and
    produced results easier to interpret physically. In particular,
    the general form of the optimal eddy viscosity obtained in this
    case was found to have little dependence on regularization,
    cf.~Figure \ref{fig:FG}b.}

  The main question left open by the results reported here is whether
  the optimal eddy viscosity for the Smagorinsky model in 3D turbulent
  flows would exhibit similar properties. It can be studied by solving
  an optimization problem analogous to \eqref{eq:minJ}, a task we will
  undertake in the near future. In addition, it is also interesting to
  analyze the optimal performance of other closure models using the
  framework developed here.

{
\appendix
\section{Gradient of the Error Functional $\J_2$ \label{sec:gradJ2}}

Here we discuss computation of the gradients
$\grad^{L^2}_{\varphi}\J_2$ and $\grad^{H^2}_{\varphi}\J_2$ of the
error functional \cref{eq:J2}. The difference with respect to the
formulation used in Section \ref{sec:adjoint} is that functional
\eqref{eq:J2} is defined in the Fourier space and we adopt with
suitable modifications the approach developed in
\cite{Farazmand2011}. Proceeding as in Section \ref{sec:adjoint}, we
first compute the G\^{a}teaux differential of the error functional
\cref{eq:J2} with respect to $\varphi$
\begin{equation} 
\J_2'(\varphi; \varphi') = {\frac{1}{2T} \int_{t=0}^{T} \int_{k = 0}^{k_c} 
  \bigg( \left[ \Ew(\cdot, k; \varphi) \right]_T -  \left[ E_w(\cdot,
    k) \right]_T  \bigg) \, \left( \int_{\mathscr{C}(k)} \htw
    \overline{\htw}' + \overline{\htw} \htw' \, dS(\kb) \right)
dk \, dt,}
\label{eq:dJ22}
\end{equation}
where $\overline{\cdot}$ denotes the complex conjugate and $\htw'$ is
the Fourier transform of the solution $\tw'$ to \cref{eq:Pert}. We
note that the gradients $\grad^{L^2}_{\varphi}\J_2$ and
$\grad^{H^2}_{\varphi}\J_2$ satisfy Riesz identities analogous to
\eqref{eq:Riesz}.  Next we introduce new adjoint fields $\tw^*$ and
$\tpsi^*$ assumed to satisfy the same adjoint system \eqref{eq:Adj},
but with a different source term $W$ whose form is to be
determined. Utilizing Parseval's identity and the fact that all fields
are real-valued in physical space, we rewrite the duality relation
\eqref{eq:dual} as
\begin{equation}
\begin{aligned}
\left( \begin{bmatrix}[1.0] \tw' \\ \tpsi' \end{bmatrix}, \K^*\begin{bmatrix}[1.0] \tw^* \\ \tpsi^* \end{bmatrix}\right) & = \frac{1}{2}\left( \widehat{\begin{bmatrix}[1.0] \tw' \\ \tpsi' \end{bmatrix}}, \widehat{\K^*\begin{bmatrix}[1.0] \tw^* \\ \tpsi^* \end{bmatrix}}\right) + \frac{1}{2}\overline{\left( \widehat{\begin{bmatrix}[1.0] \tw' \\ \tpsi' \end{bmatrix}}, \widehat{\K^*\begin{bmatrix}[1.0] \tw^* \\ \tpsi^* \end{bmatrix}}\right)}, \\
& = {\frac{1}{2T}\, \int_{t=0}^{T} \int_{k = 0}^{k_c}  \int_{\mathscr{C}(k)}
 \widehat{\left[\begin{smallmatrix} \tw' \\ \tpsi' \end{smallmatrix}\right]} \cdot \overline{\widehat{\K^*\left[\begin{smallmatrix} \tw^* \\ \tpsi^* \end{smallmatrix}\right]}} + \overline{\widehat{\left[\begin{smallmatrix} \tw' \\ \tpsi' \end{smallmatrix}\right]}} \cdot \widehat{\K^*\left[\begin{smallmatrix} \tw^* \\ \tpsi^* \end{smallmatrix}\right]} \, dS(\kb) \, dk \, dt.}
\end{aligned}
\label{eq:Parseval}
\end{equation}
Combining \eqref{eq:Pert}, \eqref{eq:dual}, \eqref{eq:Adj}, \cref{eq:dJ22} and \cref{eq:Parseval} results in
\begin{equation}
\begin{aligned}
\left( \begin{bmatrix} \tw' \\ \tpsi' \end{bmatrix}, \K^*\begin{bmatrix} \tw^* \\ \tpsi^* \end{bmatrix}\right) 
= \overbrace{{\frac{1}{2T} \int_{t=0}^{T} \int_{k = 0}^{k_c}  \bigg( \left[ \Ew(\cdot, k; \varphi) \right]_T -  \left[ E_w(\cdot, k) \right]_T  \bigg) \, \left( \int_{\mathscr{C}(k)} \htw \overline{\htw}' + \overline{\htw} \htw' \, dS(\kb) \right) dk \, dt,}}^{\J_2'({{\varphi}}; {\varphi}')} \nonumber
\end{aligned}
\end{equation}
from which we deduce the form of the source term in the adjoint system
as 
\begin{equation}
\widehat{W}(t, \kb) = \left( \left[ \Ew(\cdot, k; \varphi) \right]_T -
  \left[ E_w(\cdot, k) \right]_T  \right) \, \htw(t, \kb).
\label{eq:W2}
\end{equation}
Once the adjoint system \eqref{eq:Adj} with the source term
\eqref{eq:W2} is solved, the $L^2$ gradient
$\grad_{\varphi}^{L^2}\J_2$ can be computed using expression
\eqref{eq:gradL2}. The Sobolev gradient $\grad_{\varphi}^{H^2}\J_2$ is
then obtained as discussed in Section \ref{sec:adjoint} by solving
system \eqref{eq:gradH2BVP}. In summary, the difference in the
computation of the gradients of the error functionals $\J_1$ and
$\J_2$ is confined to the form of the source term $W$ in the adjoint
system \eqref{eq:Adj}.

}

\begin{acknowledgments}
  This research was partially supported by an NSERC (Canada) Discovery
  Grant. Computational resources were provided by Compute Canada under
  its Resource Allocation Competition. {The authors acknowledge
    helpful and constructive feedback provided by two anonymous
    referees.}

\end{acknowledgments}


%

\end{document}